\documentclass[10pt, journal,final,compsoc]{IEEEtran}

\usepackage{graphicx}
\usepackage[section]{algorithm}
\usepackage{algorithmic}
\usepackage{amsmath}
\usepackage{bm}
\usepackage{color}
\usepackage{ragged2e}
\usepackage{subfigure}
\usepackage[numbers,sort&compress]{natbib}

\renewcommand{\raggedright}{\leftskip=0pt \rightskip=0pt plus 0cm}

\ifCLASSINFOpdf
\else
\fi

\begin{document}

\title{A Parallel Patient Treatment Time Prediction Algorithm and Its Applications in Hospital Queuing-Recommendation in a Big Data Environment}

\author{Jianguo~Chen, ~\IEEEmembership{Student Member, IEEE}
        Kenli~Li, ~\IEEEmembership{Senior Member, IEEE},
        Zhuo~Tang, ~\IEEEmembership{Member, IEEE},
        Kashif~Bilal,
        and~Keqin~Li, ~\IEEEmembership{Fellow, IEEE}
\IEEEcompsocitemizethanks{\IEEEcompsocthanksitem Jianguo~Chen, Kenli~Li, Zhuo~Tang, and~Keqin~Li are with the College of Computer Science and Electronic Engineering, Hunan University, and the National Supercomputing Center in Changsha, Hunan, Changsha, 410082, China.
\protect\\
Corresponding author: Kenli Li, Email: lkl@hnu.edu.cn.

\IEEEcompsocthanksitem Kashif Bilal is with the COMSATS Institute of Information Technology, Islamabad, Pakistan, and is also with the Qatar University, Qatar.

\IEEEcompsocthanksitem Keqin Li is also with the Department of Computer Science, State University of New York, New Paltz, NY 12561, USA.
}
}

\markboth{}%
{Shell \MakeLowercase{\textit{et al.}}: Bare Advanced Demo of IEEEtran.cls for Journals}

\IEEEtitleabstractindextext{%
\renewcommand{\raggedright}{\leftskip=0pt \rightskip=0pt plus 0cm}
\begin{abstract}
 \raggedright{
Effective patient queue management to minimize patient wait delays and patient overcrowding is one of the major challenges faced by hospitals.
Unnecessary and annoying waits for long periods result in substantial human resource and time wastage and increase the frustration endured by patients.
For each patient in the queue, the total treatment time of all patients before him is the time that he must wait.
It would be convenient and preferable if the patients could receive the most efficient treatment plan and know the predicted waiting time through a mobile application that updates in real-time.
Therefore, we propose a Patient Treatment Time Prediction (PTTP) algorithm to predict the waiting time for each treatment task for a patient.
We use realistic patient data from various hospitals to obtain a patient treatment time model for each task.
Based on this large-scale, realistic dataset, the treatment time for each patient in the current queue of each task is predicted.
Based on the predicted waiting time, a Hospital Queuing-Recommendation (HQR) system is developed.
HQR calculates and predicts an efficient and convenient treatment plan recommended for the patient.
Because of the large-scale, realistic dataset and the requirement for real-time response, the PTTP algorithm and HQR system mandate efficiency and low-latency response.
We use an Apache Spark-based cloud implementation at the National Supercomputing Center in Changsha (NSCC) to achieve the aforementioned goals.
Extensive experimentation and simulation results demonstrate the effectiveness and applicability of our proposed model to recommend an effective treatment plan for patients to minimize their wait times in hospitals.}
\end{abstract}

\begin{IEEEkeywords}
Apache Spark, Big Data, Cloud Computing, Hospital Queuing Recommendation, Patient Treatment Time Prediction
\end{IEEEkeywords}
}

\maketitle
\IEEEdisplaynontitleabstractindextext
\IEEEpeerreviewmaketitle

\section{Introduction}
\subsection{Motivation}
\IEEEPARstart{C}{urrently}, most hospitals are overcrowded and lack effective patient queue management.
Patient queue management and wait time prediction form a challenging and complicated job because each patient might require different phases/operations, such as a checkup, various tests, e.g., a sugar level or blood test, X-rays or a CT scan, minor surgeries, during treatment.
We call each of these phases /operations as treatment tasks or tasks in this paper.
Each treatment task can have varying time requirements for each patient, which makes time prediction and recommendation highly complicated.
A patient is usually required to undergo examinations, inspections or tests (refereed as tasks) according to his condition.
In such a case, more than one task might be required for each patient.
Some of the tasks are independent, whereas others might have to wait for the completion of dependent tasks.
Most patients must wait for unpredictable but long periods in queues, waiting for their turn to accomplish each treatment task.

\begin{figure}[!ht]
\setlength{\abovecaptionskip}{0pt}
\setlength{\belowcaptionskip}{0pt}
\centering
\includegraphics[width=3.4in]{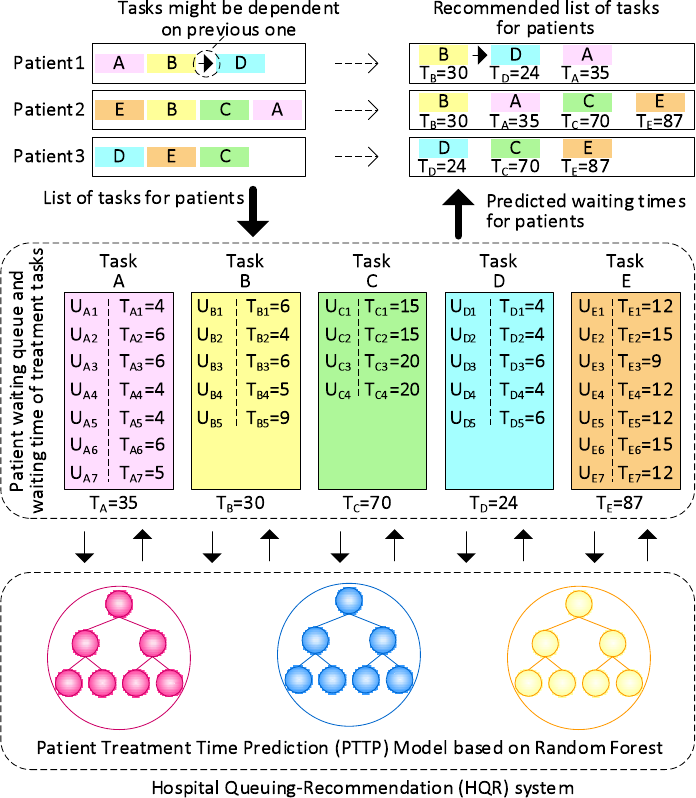}
\caption{Workflow of patient treatment and wait model}
\label{fig01}
\end{figure}

In this paper, we focus on helping patients complete their treatment tasks in a predictable time and helping hospitals schedule each treatment task queue and avoid overcrowded and ineffective queues.
We use massive realistic data from various hospitals to develop a patient treatment time consumption model.
The realistic patient data are analyzed carefully and rigorously based on important parameters, such as patient treatment start time, end time, patient age, and detail treatment content for each different task.
We identify and calculate different waiting times for different patients based on their conditions and operations performed during treatment.
The workflow of the patient treatment and wait model is illustrated in Fig. \ref{fig01}.

Fig. \ref{fig01} illustrates three patients ($Patient1$, $Patient2$, and $Patient3$) and a set of treatment tasks required for each patient.
Some tasks can be dependent on a previous one, e.g., surgery or bandage cannot be done before X-rays.
Tasks $\{A, B, D\}$ are required for $Patient1$, whereas task $D$ must wait for the completion of $B$.
Tasks $\{E, B, C, A\}$ are required for $Patient2$, and tasks $\{D, E, C\}$ are required for $Patient3$.
Moreover, there are different numbers of patients waiting in the queue of each task, for example, 7 patients in the queue of task $A$ and 5 patients in the queue of task $B$.

In this paper, a Patient Treatment Time Prediction (PTTP) model is trained based on hospitals' historical data.
The waiting time of each treatment task is predicted by PTTP, which is the sum of all patients' waiting times in the current queue.
Then, according to each patient's requested treatment tasks, a Hospital Queuing-Recommendation (HQR) system recommends an efficient and convenient treatment plan with the least waiting time for the patient.

The patient treatment time consumption of each patient in the waiting queue is estimated by the trained PTTP model.
The whole waiting time of each task at the current time can be predicted, such as
$\{T_{A} = 35(min),~ T_{B}=30(min),~ T_{C} = 70(min),~ T_{D} = 24(min),~ T_{E} = 87(min)\}$.
Finally, the tasks of each patient are sorted in an ascending order according to the waiting time, except for the dependent tasks.
A queuing recommendation is performed for each patient, such as the recommended queuing $\{ B, D, A\}$ for $Patient1$, $\{ B, A, C, E\}$ for $Patient2$, and $\{ D, C, E\}$ for $Patient3$.

To complete all of the required treatment tasks in the shortest waiting time, the waiting time of each task is predicted in real-time.
Because the waiting queue for each task updates, the queuing recommendation is recomputed in real-time.
Therefore, each patient can be advised to complete his treatment activities in the most convenient way and with the shortest waiting time.

\subsection{Our Contributions}
In this paper, we propose a PTTP algorithm and an HQR system.
Considering the real-time requirements, enormous data, and complexity of the system, we employ big data and cloud computing models for efficiency and scalability.
The PTTP algorithm is trained based on an improved Random Forest (RF) algorithm for each treatment task, and the waiting time of each task is predicted based on the trained PTTP model.
Then, HQR recommends an efficient and convenient treatment plan for each patient.
Patients can see the recommended plan and predicted waiting time in real-time using a mobile application.
Extensive experimentation and application results show that the PTTP algorithm achieves high precision and performance.

Our contributions in this paper can be summarized as follows.
\begin{itemize}
  \item A PTTP algorithm is proposed based on an improved Random Forest (RF) algorithm.
  The predicted waiting time of each treatment task is obtained by the PTTP model, which is the sum of all patients' probable treatment times in the current queue.
  \item An HQR system is proposed based on the predicted waiting time.
        A treatment recommendation with an efficient and convenient treatment plan and the least waiting time is recommended for each patient.
  \item The PTTP algorithm and HQR system are parallelized on the Apache Spark cloud platform at the National Supercomputing Center in Changsha (NSCC) to achieve the aforementioned goals.
      Extensive hospital data are stored in the Apache HBase, and a parallel solution is employed with the MapReduce and Resilient Distributed Datasets (RDD) programming model.
\end{itemize}

The remainder of the paper is organized as follows.
Section 2 reviews related work.
Section 3 details a PTTP algorithm and an HQR system.
The parallel implementation of the PTTP algorithm and HQR system on the Apache Spark cloud environment is detailed in Section 4.
Experimental results and evaluations are presented in Section 5 with respect to the recommendation accuracy and performance.
Finally, Section 6 concludes the paper with future work and directions.

\section{Related Work}
To improve the accuracy of the data analysis with continuous features, various optimization methods of classification and regression algorithms are proposed.
A self-adaptive induction algorithm for the incremental construction of binary regression trees was presented in \cite{ex01}.
Tyree et al. \cite{ex02} introduced a parallel boosted regression tree algorithm for web search ranking.
In \cite{ex03}, a multi-branch decision tree algorithm was proposed based on a correlation-splitting criterion.
Other improved classification and regression tree methods were proposed in \cite{ex04,ex05,ex06}.

The random forest algorithm \cite{ex07} is an ensemble classifier algorithm based on a decision tree, which is a suitable data-mining algorithm for big data.
The random forest algorithm is widely used in many fields such as fast action detection via discriminative random forest voting and Top-K subvolume search\cite{ex09}, robust and accurate shape model matching using random forest regression voting\cite{ex10}, and a big data analytic framework for peer-to-peer botnet detection using random forests\cite{ex11}.
The experimental results in these papers demonstrate the effectiveness and applicability of the random forest algorithm.
Bernard \cite{ex12} proposed a dynamic training method to improve the accuracy of the random forest algorithm.
In \cite{ex13}, a random forest method based on weighted trees was proposed to classify high-dimensional noisy data.
However, the original random forest algorithm uses a traditional direct voting method in the voting process.
In such a case, the random forest containing noisy decision trees would likely lead to an incorrect predicted value for the testing dataset \cite{ex08}.

Various recommendation algorithms have been presented and applied in related fields.
Meng et al.\cite{ex14} proposed a keyword-aware service recommendation method on MapReduce for big data applications.
A travel recommendation algorithm that mines people's attributes and travel-group types was proposed in \cite{ex15}.
Zu al. \cite{ex16} introduced a Bayesian-inference-based recommendation system for online social networks, in which a user propagates a content rating query along the social network to his direct and indirect friends.
Adomavicius et al. \cite{ex17} introduced new recommendation techniques for multi-criteria rating systems.
Gediminas et al. \cite{ex18} introduced an overview of the current generation of recommendation methods, such as content-based, collaborative, and hybrid recommendation approaches.
However, there is no effective prediction algorithm for patient treatment time consumption in the existing studies.

The speed of data mining and analysis for big data is a very important factor \cite{ex19}.
Cloud computing, distributed computing, and supercomputers offer high-speed computing power.
Both the Apache Hadoop \cite{ex20} and Spark \cite{ex21} are famous cloud platforms that are widely used in parallel computing and data analysis.
Numerous parallel data-mining algorithms have been implemented based on the MapReduce \cite{ex22} and RDD \cite{ex23} models.
In \cite{ex24, ex25, ex26, ex27}, various data-mining algorithms were proposed based on the MapReduce programming model.
Apache Spark is an efficient cloud platform that is suitable for data mining and machine learning.
In the Spark, data are cached in memory, and iterations for the same data come directly from memory.
Zaharia \cite{ex28} presented a fast and interactive analytics over Hadoop data with Spark.

To predict the waiting time for each treatment task, we use the random forest algorithm to train the patient treatment time consumption based on both patient and time characteristics and then build the PTTP model.
Because patient treatment time consumption is a continuous variable, a Classification And Regression Tree (CART) model is used as a meta-classifier in the RF algorithm.
Because of the shortcomings of the original RF algorithm and the characteristics of the patient data, in this paper, the RF algorithm is improved in 4 aspects to obtain an effective result from large-scale, high dimensional, continuous, and noisy patient data.
Compared with the original RF algorithm, our PTTP algorithm based on an improved RF algorithm has significant advantages in terms of accuracy and performance.
Moreover, there is no existing research on hospital queuing management and recommendations.
Therefore, we propose an HQR system based on the PTTP model.
To the best of our knowledge, this paper is the first attempt to solve the problem of patient waiting time for hospital queuing service computing.
A treatment queuing recommendation with an efficient and convenient treatment plan and the least waiting time is recommended for each patient.

\section{Patient Treatment Time Prediction Algorithm}
To build the PTTP model based on both patient and time characteristics, a PTTP algorithm is proposed.
The PTTP model is based on an improved RF algorithm and is trained from the massive, complex, and noisy hospital treatment data.

\subsection{Problem Definition and Data Preprocessing}
\subsubsection{Problem Definition}
Prediction based on analysis and processing of massive noisy patient data from various hospitals is a challenging task.
Some of the major challenges are the following:

(1) Most of the data in hospitals are massive, unstructured, and high dimensional.
Hospitals produce a huge amount of business data every day that contain a great deal of information, such as patient information, medical activity information, time, treatment department, and detailed information of the treatment task.
Moreover, because of the manual operation and various unexpected events during treatments, a large amount of incomplete or inconsistent data appears, such as a lack of patient gender and age data, time inconsistencies caused by the time zone settings of medical machines from different manufacturers, and treatment records with only a start time but no end time.

(2) The time consumption of the treatment tasks in each department might not lie in the same range, which can vary according to the content of tasks and various circumstances, different periods, and different conditions of patients.
For example, in the case of a CT scan task, the time required for an old man is generally longer than that required for a young man.

(3) There are strict time requirements for hospital queuing management and recommendation.
The speed of executing the PTTP model and HQR scheme is also critical.

\subsubsection{Data Preprocessing}
In the preprocessing phase, hospital treatment data from different treatment tasks are gathered.
Substantial numbers of patients visit each hospital every day. Let $S$ be a set of patients in a hospital, and a patient who has been registered and his information is represented by $s_{i}$. Assume that there are $N$ patients in $S$:
\begin{center}
$S = \{s_{1},~ s_{2},~ ...,~ s_{N}\}$, \\
\end{center}
where each patient $s_{i}$ can have specific unchanged parameters, e.g., name, ID, gender, age, and address. Some of these parameters are useful to our analysis, whereas others are not.

Each patient can visit multiple treatment tasks according to his health condition.
Let $X|s_{i}$ be a set of treatment tasks for patient $s_{i}$ during a specific visit:
\begin{center}
$X|s_{i} = \{x_{1},~ x_{2},~ ...,~ x_{K}\}$,\\
\end{center}
where each treatment task record $x_{i}$ can consist of multiple information $Y$, e.g., task name, task location, department, start time, end time, doctor, and attending staff:
\begin{center}
$Y|x_{i} = \{y_{1},~ y_{2},~ ...,~ y_{M}\}$,
\end{center}
where $y_{j}$ is a feature variable of the record of treatment task $x_{i}$.
Here, for a single visit, we have a single record for patient name, age, gender, and multiple records for treatment tasks, as shown in Table \ref{table310}.

\begin{table}[htbp]
\setlength{\abovecaptionskip}{0pt}
\setlength{\belowcaptionskip}{0pt}
\renewcommand{\arraystretch}{1.0}
\caption{Example of treatment records}
\label{table310}
\begin{tabular*}{3.5in}{p{0.2in} p{0.2in} p{0.08in} p{0.3in} p{0.3in} p{0.2in} p{0.52in} p{0.52in}}
\hline
Patient  & Gen & Age & Task & Dept.  & Doctor  & Start  & End \\
No. & ~ & ~ &  name &  name &  name &  time &  time\\
\hline
0001 & Male & 15 & Checkup & Surgery & Dr. Chen & 2015-10-10 08:30:00 & 2015-10-10 08:42:25 \\

0001 & Male & 15 & Payment & Cashier-6 & \emph{Null} & 2015-10-10 08:50:05 & \emph{Null} \\

0001 & Male & 15 & CT scan & CT-5 & Dr. Li & 2015-10-10 09:20:00 & 2015-10-10 09:27:00 \\

0001 & Male & 15 & MR scan & MR-8 & Dr. Pan & 2015-10-10 10:05:06 & 2015-10-10 10:15:35 \\

0001 & Male & 15 & Take medicine & TCM Pharmacy & \emph{Null} & 2015-10-10 10:42:03 & 2015-10-10 10:45:29 \\

... & ... & ... & ... & ... & ... & ... & ... \\
\hline
\end{tabular*}
\end{table}

The workflow of the preprocessing task can be depicted by the following steps.

\textbf{(1) Gather data from different treatment tasks.}

Depending on statistics, the number of patients in a medium-sized hospital lies between 8,000 and 12,000 per day, and the number of remedial treatment data records is between 120,000 and 200,000.
These data are gathered from different treatment tasks, including registration, medical examination, inspection, drug delivery, payment, and other related tasks.
The formats of the data for different treatment tasks are shown in Table \ref{table30}.

\begin{table}[htbp]
\setlength{\abovecaptionskip}{0pt}
\setlength{\belowcaptionskip}{0pt}
\renewcommand{\arraystretch}{1.2}
\caption{Formats of the data for different treatment tasks}
\label{table30}
\tabcolsep1pt
\begin{tabular*}{3.5in}{c p{2.7in}l}
\hline
Treatment task & Format of the data (Feature name)\\
\hline
Registration & \{Patient card number, patient name, gender, age, telephone number, address, task name, operation time\} \\

Checkup & \{Patient card number, patient name, gender, age, task name, department, doctor name, doctor position, start time, end time, context\}\\

Payment & \{Patient card number, patient name, task name, amount, operation time\}\\

Take medicine & \{Patient card number, patient name, task name, dispensary, time of compounding, time of issue\}\\

CT scan & \{Patient card number, patient name, gender, age, task name, department, doctor, body region of scans, start time, end time, remark\}\\

Injection & \{Patient card number, patient name, gender, age, task name, department, doctor, start time, end time, drug name, drug number, remark \}\\

Blood Tests& \{Patient card number, patient name, gender, age, task name, department, doctor, time of blood tests, time of report\}\\

... & ...\\
\hline
\end{tabular*}
\end{table}

\textbf{(2) Choose the same dimensions of the data.}

The hospital treatment data generated from different treatment tasks have different contents and formats as well as varying dimensions.
To train the patient time consumption model for each treatment task, we choose the same features of these data, such as the patient information (patient card number, gender, age, etc.), the treatment task information (task name, department name, doctor name, etc.), and the time information (start time and end time).
Other feature subspaces of the treatment data are not chosen because they are not useful for the PTTP algorithm, such as patient name, telephone number, and address.

\textbf{(3) Calculate new feature variables of the data.}

To train the PTTP model, various important features of the data should be calculated, such as the patient time consumption of each treatment record, day of week for the treatment time, and the time range of treatment time.
For example, in the treatment record of the CT scan task in Table \ref{table310}, the start time is ``2015-10-10 09:20:00" and the end time is ``2015-10-10 09:27:00", the time consumption for this patient in the treatment is ``420 (s)", the day of the week is ``Saturday", and the time range is ``09".

\textbf{(4) Remove incomplete and inconsistent data.}

After calculating new feature variables of treatment data, the error and noisy data need to be removed.
The treatment records with missing values for critical features are removed as incomplete data, such as patient gender, patient age, and task name.
The treatment records with negative values of time consumption are removed as inconsistent data, for instance, if the end time of the treatment operation is before the start time, which can occur in cases when a start time is recorded by a human and an end time is shown by a machine.
The types of data shown above are considered as noisy data in this paper.
The features of the treatment data used in the process of employing the PTTP algorithm are presented in Table \ref{table31}.

\begin{table}[!ht]
\setlength{\abovecaptionskip}{0pt}
\setlength{\belowcaptionskip}{0pt}
\renewcommand{\arraystretch}{1.2}
\caption{Features of treatment data for the PTTP algorithm}
\label{table31}
\tabcolsep1pt
\begin{tabular*}{3.5in}{c c p{2.2in}l}
\hline
No. & Feature Name & Value range of each feature subspace\\
\hline
 $y_{1}$ & Patient Gender & ``Male", ``Female".\\

 $y_{2}$ & Patient Age & The age of the patient.\\

 $y_{3}$ & Department & All departments in the hospital.\\

 $y_{4}$ & Doctor Name & All doctors in the hospital.\\

 $y_{5}$ & Task Name & Each treatment task in all treatment processes in the hospital.\\

 $y_{6}$ & Start Time & The start time of the treatment task.\\

 $y_{7}$ & End Time & The end time of the treatment task.\\

 $y_{8}$ & Week & The day of week for the treatment time. The value is from Monday to Sunday.\\

 $y_{9}$ & Time Range & The time range of treatment time in a day. The value is from 0 to 23.\\

 $y_{10}$ & Time Consumption & (1) End time - Start time, such as a CT scan, an MR scan.
(2) Time interval between one patient and the next in the same treatment, such as payment.\\
\hline
\end{tabular*}
\end{table}

\subsubsection{Constructing Training Subsets for the PTTP Model}
In the process of employing the PTTP model, the treatment time consumption of patients with different conditions and different environments in each treatment task are addressed.
Due to the diverse nature of different medical tasks, the range of patient treatment time consumption cannot be measured by an absolute standard.

To improve the accuracy of the PTTP model, an improved RF algorithm is used to build the PTTP model.
$k$ training subsets are sampled from the original training dataset $S$ in a bootstrap sampling process.
$N$ samples are selected from $S$ by a random sampling and replacement method in each sampling period.
After the current step, $k$ training subsets are constructed as a collection of $S_{Train}$:

\begin{center}
$S_{Train} = \{s_{train1},~ s_{train2},~ ...,~ s_{traink}\}$.
\end{center}

\begin{figure}[!ht]
\setlength{\abovecaptionskip}{0pt}
\setlength{\belowcaptionskip}{0pt}
\centering
\includegraphics[width=2.8in]{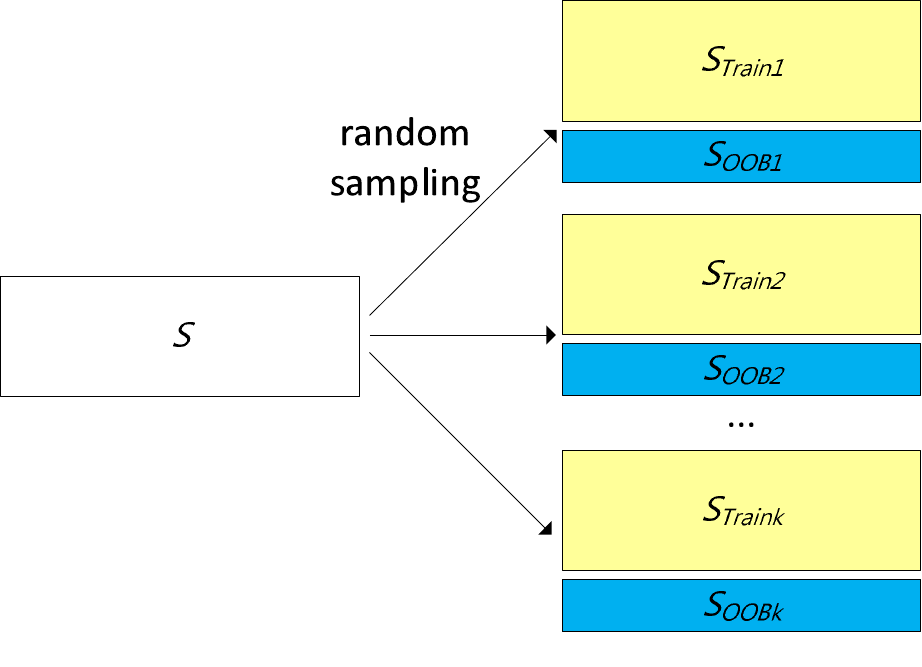}
\caption{Process of training dataset random sampling for the PTTP model}
\label{fig02}
\end{figure}

At the same time, the unselected data in each sampling period are composed as an out-of-bag (OOB) dataset.
$k$ OOB sets are constructed as a collection of $S_{OOB}$:

\begin{center}
$S_{OOB} = \{S_{OOB1},~ S_{OOB2},~ ...,~ S_{OOBk}\}$,
\end{center}
where $k \ll N$, $S_{Train}\in S$, and $S_{OOB}\in S$.
These datasets are used as testing sets after the training process to verify the classification or regression accuracy of each tree.
The process of the training dataset random sampling for the RF model is shown in Fig. \ref{fig02}.

\subsection{PTTP Model based on the improved RF Algorithm}
To predict the waiting time for each patient treatment task, the patient treatment time consumption based on different patient characteristics and time characteristics must first be calculated.
The time consumption of each treatment task might not lie in same range, which varies according to the content of tasks and various circumstances, different periods, and different conditions of patients.
Therefore, we use the RF algorithm to train patient treatment time consumption based on both patient and time characteristics and then build the PTTP model.

Because of the limitations of the original RF algorithm and the characteristics of hospital treatment data, the RF algorithm is improved in 4 aspects to obtain an effective result from large-scale, high dimensional, continuous, and noisy hospital treatment data.

(1) All of the selected (cleaned) features of the data are used in the training process, instead of $m$ features selected randomly, as is done in the original RF algorithm, because the features of the data are limited and the data are already cleaned of unnecessary features such as patient name, address, and telephone number.

(2) Because the target variable of the treatment data is patient treatment time consumption, which is a continuous variable, a CART model is used as a meta-classifier in the improved RF algorithm.
At the same time, some independent variables of the data are nominal data, which have different values such as time range (0 - 23) and day of week (Monday - Sunday).
In such a case, the two-fork tree model of the traditional CART cannot fully reflect the analysis results.
Therefore, to construct the regression tree model felicitously, a multi-branch model is proposed for the construction process instead of the two-fork model of the traditional CART algorithm.

(3) Although we have removed part of the error in the preprocessing, other types of noisy data might also exist.
In some treatment tasks, the time consumption is the time interval between one patient and the next in the same treatment.
For example, in a payment task, assume that the operation time point of the last patient in the morning is ``12:00:00" and the operation time point of the first patient in the afternoon is ``14:00:00".
The time consumption of the former is ``7200 (s)" and is considered as incorrect data because it is larger than the normal value of ``100 (s)".
However, the value ``7200 (s)" of time consumption has not always been incorrect data, such as in a blood examination task.
Therefore, we cannot simply designate one value of time consumption as noisy data; each must be classified according to treatment data features.
Then, we must identify and remove the noisy data.
In calculating the average value of the data in each leaf node of the regression tree, noisy data are removed to reduce their influence on accuracy.

(4) The original RF algorithm uses a traditional direct voting method in the prediction process.
In such a case, a RF containing noisy decision trees would likely lead to an incorrect predicted value for the testing dataset.
Therefore, in this paper, a weighted voting method is employed in the prediction process of the RF model.
Each tree classifier corresponds to a specified reasonable weight for voting the testing data.
A tree classifier that has high accuracy in the training process will have a high voting weight in the prediction process.
Hence, the classifier improves the overall classification accuracy of the RF algorithm, and reduces the generalization error.

Compared with the original RF algorithm, our PTTP algorithm based on the improved RF algorithm, has significant advantages in terms of accuracy and performance.

\subsubsection{Training CART Regression Trees of the RF Model}
Because the patient treatment time consumption is the target feature variable of treatment data $S$, which is a continuous value, the type of the single decision tree in the RF model is a regression tree.
Thus, a CART regression tree model is created for each training subset $s_{traini}$.

The first optimization aspect of the RF algorithm is in the growing process of each CART tree.
All of the $M$ features of each training data $s_{traini}$ are used in the training process instead of the $m$ features selected randomly as is done in the original RF algorithm.
The main process of building the regression tree of CART is described as follows.

\textbf{(1) Calculate the best splitting feature variables and the best split point.}

In each tree node's splitting process, each feature variable subspace $y_{j}$ and each potential split point value $v_{p}$ of $y_{j}$ are chosen to calculate the loss function of $(y_{j}, v_{p})$, which is defined as follows:

\begin{equation}
\label{equ01}
\begin{aligned}
(y_{j}, v_{p}) =& arg \min[\sum_{x \in R_{L}(y_{j}, v_{p})}(y_{i}-c_{L})^{2} \\
                & + \sum_{x \in R_{R}(y_{j}, v_{p})}(y_{i}-c_{R})^{2}],
\end{aligned}
\end{equation}
where a summary of the elements in Eq. (\ref{equ01}) is presented in Table \ref{table32}.

\begin{table}[!ht]
\setlength{\abovecaptionskip}{0pt}
\setlength{\belowcaptionskip}{0pt}
\renewcommand{\arraystretch}{1.3}
\caption{Summary of the elements in Eq. (\ref{equ01})}
\label{table32}
\tabcolsep1pt
\begin{tabular*}{3.5in}{c p{2.4in}l}
\hline
 Element & Description\\
\hline
 $y_{j}$ & each feature subspace of the training dataset, $1\leq j \leq M$.\\

 $v_{p}$ & each potential split point value of $y_{j}$. \\

 $R_{L}(y_{j},~ v_{p})$ & the first (left) subset of data split by $v_{p}$ in the feature subspace $y_{j}$.\\

 $R_{R}(y_{j},~ v_{p})$ & the second (right) subset of data split by $v_{p}$ in the feature subspace $y_{j}$.\\

 $c_{L}$ & the average value in the $R_{L}(y_{j},~ v_{p})$ subset.\\

 $c_{R}$ & the average value in the $R_{R}(y_{j},~ v_{p})$ subset.\\
\hline
\end{tabular*}
\end{table}

In such a case, the variable $y_{j}$ with the smallest value of the loss function is selected as the best split feature, and the value $v_{p}$ is used as the split point for $y_{j}$ at the current splitting tree node.

\textbf{(2) Split the data into two forks.}

Split the training dataset into two forks by $v_{p}$ in the feature subspace $y_{j}$.
$R_{L(y_{j},~ v_{p})}$ denotes the first (left) data subset and $R_{R(y_{j},~ v_{p})}$ denotes the second (right) data subset.
These subsets are defined as follows:

\begin{equation}
\label{equ02}
\begin{aligned}
R_{L(y_{j},~ v_{p})} = \{x | (y_{j} \leq v_{p})\}, \\
R_{R(y_{j},~ v_{p})} = \{x | (y_{j} > v_{p})\}.
\end{aligned}
\end{equation}

\textbf{(3) Construct multi-branch for the CART model.}

Some independent variables of data are nominal data, which have different values, such as the time range (0 - 23) and day of week (Monday - Sunday).
Therefore, to construct the regression tree model felicitously, a multi-branch regression tree model instead of two-fork tree model is used constructing the CART, which is the second optimization aspect of the RF algorithm.
After the tree node split into two forks by variable $y_{j}$ and value $v_{p}$ in step (2), the same variable $y_{j}$ continues to be selected to calculate the best split point $v_{pL}$ for the data in the left branch and $v_{pR}$ for the data in the right branch.
Taking the left branch as an example, the best split point calculated for the current feature subspace is defined as follows:

\begin{equation}
\label{equ03}
\Phi(v_{pL}|y_{j})=\max_{i} \Phi(v_{i}|y).
\end{equation}

The $\Phi(v_{i}|y)$ is defined as follows:
\begin{equation}
\label{equ04}
\Phi(v_{i}|y)=2P_{L}P_{R}\sum_{j=1}^{m}|p(c_{j}|y_{L}) - p(c_{j}|y_{R})|,
\end{equation}
where $P_{L}$ and $P_{R}$ are the ratios of the amount of data in the left branch and in the right branch to the entire volume of training data, respectively.
$p(c_{j}|y_{L})$ is the ratio of the volume of data that belong to class $c_{j}$ in the left branch to the volume of data in the left branch.

\begin{figure}[!ht]
\setlength{\abovecaptionskip}{0pt}
\setlength{\belowcaptionskip}{0pt}
\centering
\includegraphics[width=3.5in]{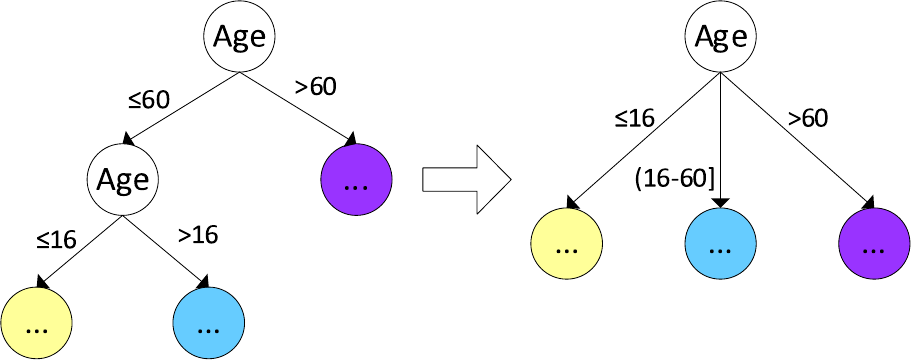}
\caption{Example of multi-branch splitting for the CART model}
\label{fig03}
\end{figure}

If the split value of $\Phi(v_{pL}|y_{j})$ is greater than the father node, namely $\Phi(v_{pL}|y_{j}) \geq \Phi(v_{p}|y_{j})$, then the left branch continues to split by the variable $y_{j}$ and value $v_{pL}$.
Otherwise, the remaining feature variables continue to be computed.
The right branch is calculated similarly.
Then, each node and its two subnodes are calculated successively.
If the same variable split exists in both the parent node and the child node, a node merger operation should be done.
Consequently, a multi-branch node of the tree is constructed.
An example of multi-branch splitting for the CART model is shown in Fig. \ref{fig03}.

Repeat steps (1 - 3) until the data in each branch are classified in one class as a leaf node.

\textbf{(4) Calculate mean value of leaf nodes after removal of noisy data.}

Although we have removed part of the error data in the preprocessing, other types of noisy data mentioned above might exist.
Therefore, the third optimization aspect of the RF algorithm is to reduce the influence that the noisy data have on the algorithm accuracy. A box-plot-based noise removal method is performed in the value calculation of each CART leaf node.

The data in the current leaf node are sorted in ascending order.
Then, the values of three data points $Q1$, $Q2$, $Q3$ of the box-plot model are calculated, where $Q2$ is the median data point and $Q1$ and $Q3$ are the lower and upper four digits of the data, respectively.
The inner limit of the noisy data is defined as follows:
\begin{equation}
\label{equ05}
IL = Q1 - 1.5(IQR) = Q1 - 1.5(Q3-Q1).
\end{equation}

The outer limit of the noisy data is defined as follows:
\begin{equation}
\label{equ06}
OL = Q3 + 1.5(IQR) = Q3 + 1.5(Q3-Q1).
\end{equation}

The data outside the range of $\{IL, OL\}$ are removed as noisy data.
After removing the noisy data, the average value $ c_{j}$ of the data $y_{j}$ is calculated in each leaf node of the regression tree.
The calculation formula is defined as follows:
\begin{equation}
\label{equ07}
c_{j} = \frac{1}{k} \sum{y_{j}},~~~~   (IL \leq y_{j} \leq OL),
\end{equation}
where $k$ is the number of data items in the current leaf node.

\begin{figure}[!ht]
\setlength{\abovecaptionskip}{0pt}
\setlength{\belowcaptionskip}{0pt}
\centering
\includegraphics[width=3.5in]{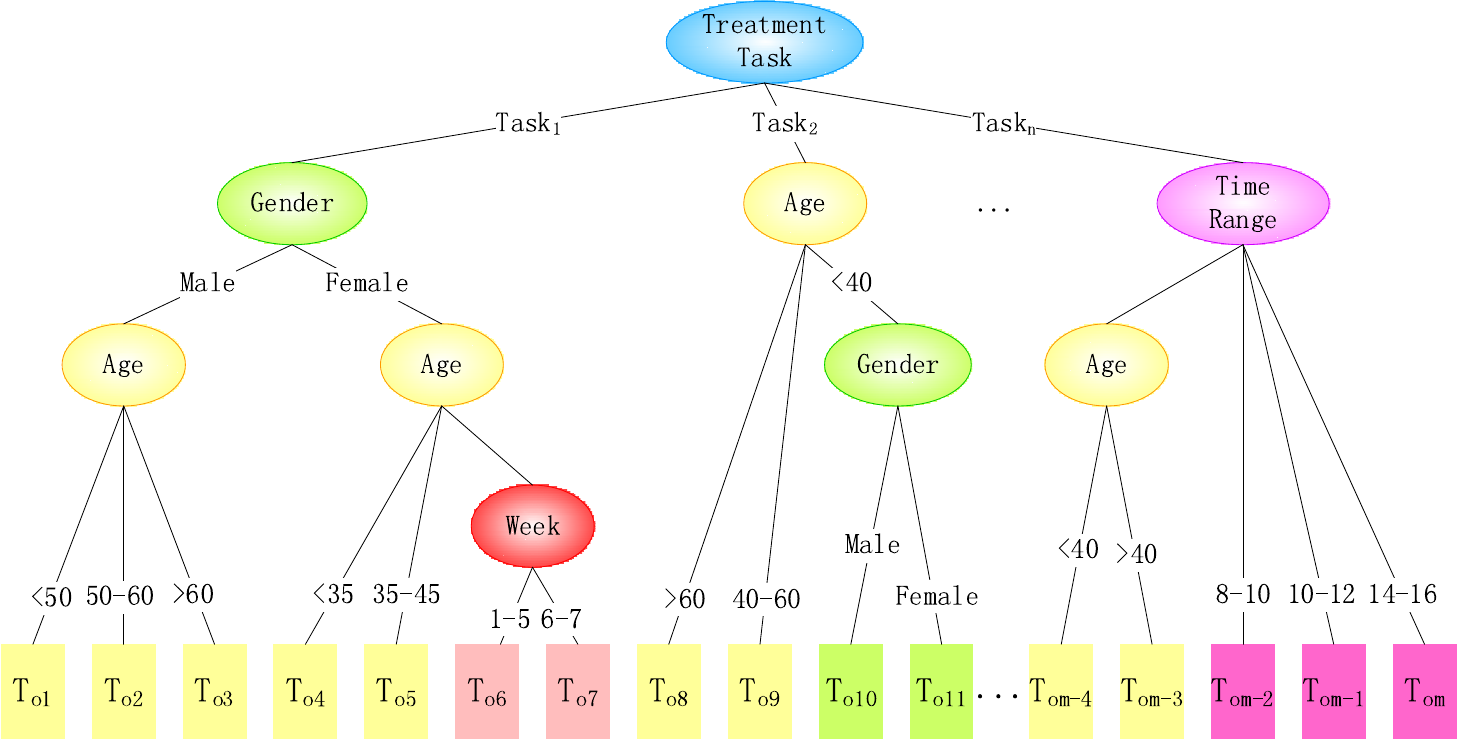}
\caption{Meta CART tree of the PTTP model}
\label{fig04}
\end{figure}

This splitting process is repeated until all of the feature values are generated.
A CART regression tree for the training subset $S_{traini}$ is trained, and the tree model is defined as follows:
\begin{equation}
\label{equ08}
h_{i}(x,~ \Theta j) = \sum_{n=1}^{N}{c_{n}I(x\in R_{n})},
\end{equation}
where $N$ is the number of leaf nodes of the tree, $\Theta j$ is the target feature variable, and $I(\cdot)$ is an indicator function.
A meta CART regression tree of the PTTP model is shown in Fig. \ref{fig04}.

\textbf{(5) Calculate the accuracy of each tree.}

After each regression tree of the training subset $S_{traini}$ is built, the testing subset $S_{OOBi}$ is used to calculate the accuracy of the meta-classifier tree.
The accuracy of a meta-classifier tree refers to the ratio of average number of votes in correct classes to all of the error classes, which are classified by the trained meta-classifier tree.
The accuracy of each meta CART tree $h_{i}(x)$ is defined as follows:

\begin{equation}
\label{equ09}
CA_{i}=\frac{I(h_{i}(x,~ \Theta j)=y)}{I(h_{i}(x,~ \Theta j)=y)+ \sum{I(h_{i}(x,~ \Theta j)=z)}},
\end{equation}
where $y$ is a value in the correct class, and $z$ is a value in the error class ($z\neq y$).

\subsubsection{Collecting $k$ CART Trees for a RF Model}

\begin{figure}[!ht]
\setlength{\abovecaptionskip}{0pt}
\setlength{\belowcaptionskip}{0pt}
\centering
\includegraphics[height = 1.7in]{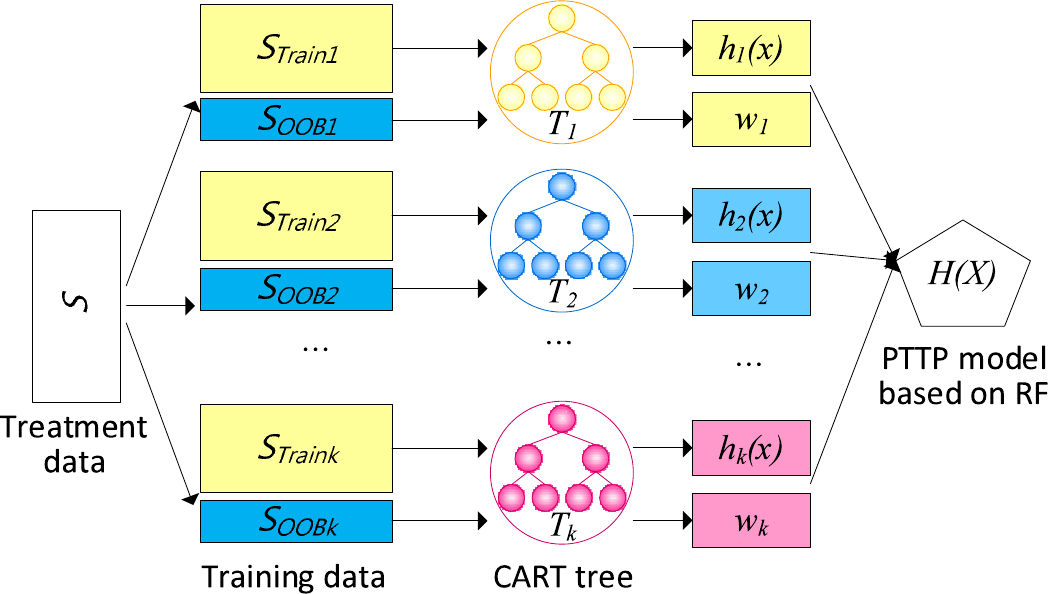}
\caption{PTTP model based on the RF algorithm}
\label{fig05}
\end{figure}

After the construction of the $k$ CART regression trees, these trees are collected for a random forest model.
A method of weighted average addition is used for the RF model, which is the fourth optimization aspect for the RF algorithm.
The weighted regression result $H(X)$ of the RF model for the data $X$ is the average value of $k$ trees, which is defined as follows:

\begin{equation}
\label{equ10}
\begin{aligned}
H(X,~ \Theta j) &=\frac{1}{k} \sum_{i=1}^{k} {[w_{i}\times h_{i}(x,~ \Theta j)]}\\
&=\frac{1}{k}\sum_{i=1}^{k}{[CA_{i}\times h_{i}(x,~ \Theta j)]},
\end{aligned}
\end{equation}
where $w_{i}$ is the weight of tree $h_{i}$ and $h_{i}(x,~ \Theta j)$ is a meta-classifier for a pruning regression tree constructed by the CART algorithm.
The PTTP model based on the random forest algorithm is shown in Fig. \ref{fig05}.

The detailed steps of the PTTP model based on the random forest algorithm are presented in Algorithm \ref{alg31}.

\begin{algorithm}[!ht]
\caption{Process of the RF-based PTTP algorithm}
\label{alg31}
\begin{algorithmic}[1]
\REQUIRE ~\\
    $S_{Train}$: the training datasets;\\
    $k$: the number of CART trees in the RF model.
\ENSURE ~\\
    $PTTP_{RF}$: the PTTP model based on the RF algorithm.
\FOR{$i = 1$ to $k$}
\STATE create training subset $s_{traini} \leftarrow sampling(S_{Train})$;
\STATE create OOB subset $s_{OOBi} \leftarrow (S_{Train} - s_{traini})$;
\STATE create an empty CART tree $h_{i}$;
\FOR {each independent variable $y_{j}$ in $s_{traini}$}
\STATE calculate candidate split points  $v_{s} \leftarrow y_{j}$;
\FOR {each $v_{p}$ in $v_{s}$}
\STATE calculate the best split point $(y_{j}, v_{p}) \leftarrow arg\min[\sum_{x \in R_{L}}(y_{i}-c_{L})^{2} + \sum_{x \in R_{R}}(y_{i}-c_{R})^{2}]$;
\ENDFOR
\STATE append node $Node_{(y_{j},~ v_{p})}$ to $h_{i}$;
\STATE split data for left branch $R_{L(y_{j}, v_{p})}\leftarrow\{x|y_{j}\leq v_{p}\}$;
        \STATE split data for right branch $R_{R(y_{j}, v_{p})}\leftarrow\{x|y_{j}> v_{p}\}$;
        \FOR{ each data $R$ in $\{R_{L(y_{j}, v_{p})},~ R_{R(y_{j}, v_{p})} \}$}
            \STATE calculate $\Phi(v_{pL}|y_{j}) \leftarrow \max_{i} \Phi(v_{i}|y)$;
            \IF {($\Phi(v_{p(L|R)}|y_{j}) \geq \Phi(v_{p}|y_{j})$)}
            \STATE append subnode $Node_{(y_{j}, v_{p(L|R)})}$ to $Node_{(y_{j}, v_{p})}$ as multi-branch;
            \STATE split data to two forks $R_{L(y_{j}, v_{pL})}$ and $R_{R(y_{j}, v_{pR})}$;
            \ELSE
            \STATE collect cleaned data for leaf node $D_{leaf} \leftarrow (IL \leq y_{j} \leq OL)$;
            \STATE calculate mean value of leaf node $c \leftarrow \frac{1}{k} \sum{D_{leaf}}$;
            \ENDIF
        \ENDFOR
        \STATE remove $y_{j}$ from $s_{traini}$;
        \ENDFOR
        \STATE calculate accuracy $CA_{i} \leftarrow \frac{I(h_{i}(x)=y)}{I(h_{i}(x)=y)+ \sum{I(h_{i}(x)=z)}}$ for $h_{i}$ by testing $s_{OOBi}$;
    \ENDFOR
\STATE $PTTP_{RF} \leftarrow H(X, \Theta j) \leftarrow\frac{1}{k}\sum_{i=1}^{k}{[CA_{i}\times h_{i}]}$;
\RETURN $PTTP_{RF}$.
\end{algorithmic}
\end{algorithm}

\subsection{Hospital Queuing Recommendation System based on PTTP Model}
After training the PTTP model for each treatment task using historical hospital treatment data, a PTTP-based hospital queue recommendation system is developed.
An efficient and convenient treatment plan is created and recommended to each patient to achieve intelligent triage.

\begin{figure}[!ht]
\setlength{\abovecaptionskip}{0pt}
\setlength{\belowcaptionskip}{0pt}
\centering
\includegraphics[width=3.5in]{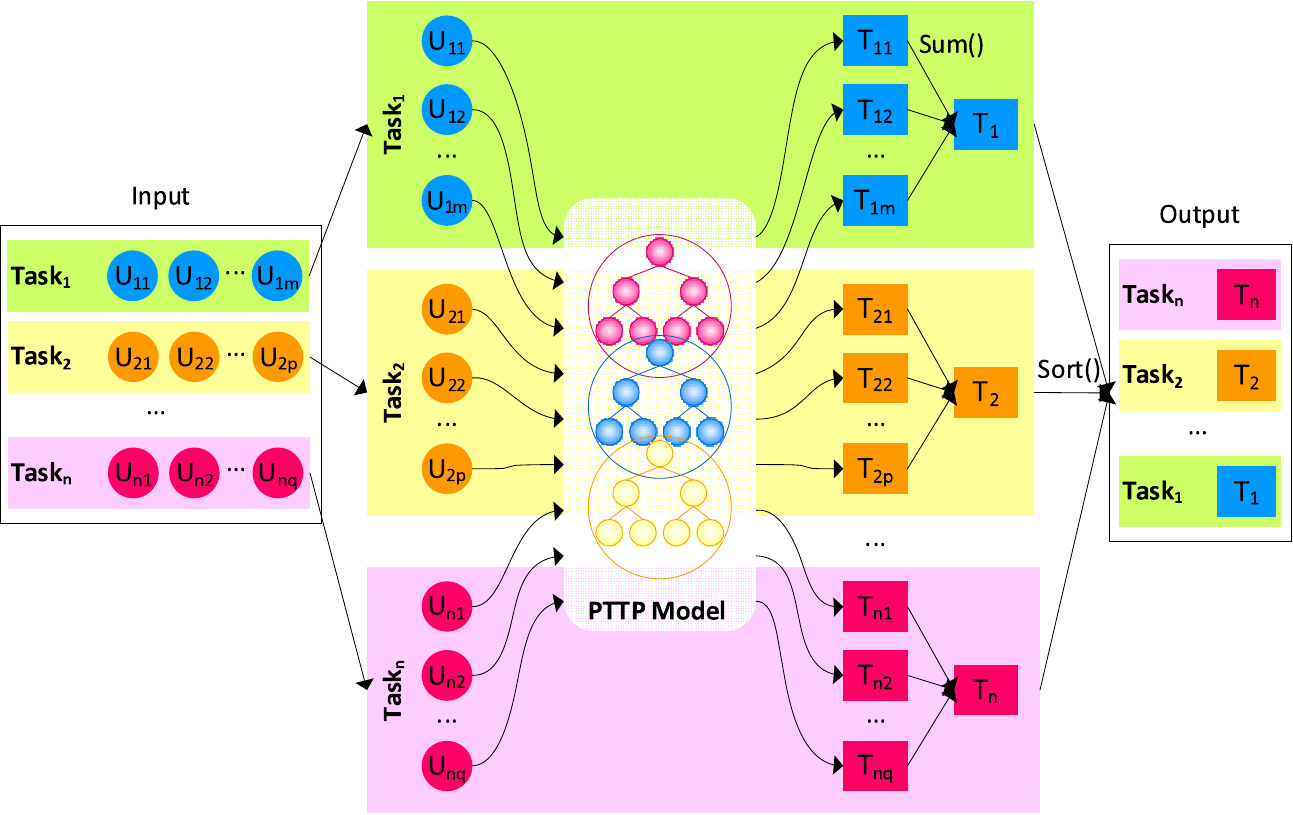}
\caption{Process of the HQR system based on the PTTP model}
\label{fig06}
\end{figure}

Assume that there are various treatment tasks for each patient according to the patient's condition, such as examinations and inspections.
Let $Tasks = \{Task_{1},~ Task_{2},~ ...,~ Task_{n}\}$ be a set of treatment tasks that the current patient must complete, and let $U_{i} = \{U_{i1},~ U_{i2},~ ...,~ U_{im}\}$ be a set of patients in waiting the queue for $Task_{i}$.
The process of the HQR system based on the PTTP model is shown in Fig. \ref{fig06}.

\textbf{(1) Predict the waiting time of all of the treatment tasks for the current patient.}

For each patient $U_{ik}$ waiting in the queue of $Task_{i}$, the patient treatment time consumption is predicted by the trained PTTP model according to the patient's characteristics (such as gender and age), time factors (such as the week and month of the current time), and other factors (such as treatment departments, available machines, and service windows).
The patient treatment time consumption $T_{ik}$ of patient $U_{ik}$ in queue is defined as follows:

\begin{equation}
\label{equ11}
\begin{aligned}
T_{ik} &=H(X_{ik},~ \Theta j) \\
 &=\frac{1}{k} \sum_{i=1}^{k} {[CA_{i}\times h_{i}(x,~ \Theta j)]},
\end{aligned}
\end{equation}
where $X_{ik}$ is the treatment data of patient $U_{ik}$, $\Theta j$ is all of the independent variables of $X_{ik}$, $CA_{i}$ is the accuracy weight of tree $h_{i}$, and $h_{i}(x,~ \Theta j)$ is a result of patient treatment time consumption predicted by a single CART regression tree.

Then, all of the predicted patient treatment time consumption of patients in the queue is added to obtain the waiting time of $Task_{i}$, which is defined as $T_{i}$.
The calculation formula of $T_{i}$ is defined as follows:

\begin{equation}
\label{equ12}
\begin{aligned}
T_{i} =\frac{1}{W_{i}} \sum_{k=1}^{m}{T_{ik}},
\end{aligned}
\end{equation}
where $W_{i}$ is the number of service windows or workbenches that can provide a service for treatment task $Task_{i}$ in parallel, $m$ is the number of patients waiting in the queue of $Task_{i}$ , and $T_{ik}$ denotes the predicted waiting time for the patient-in-waiting $Patient_{k}$.

\textbf{(2) Sort all of the treatment tasks of the current patient in ascending order by waiting time.}

All treatment tasks of the current patient are sorted in ascending order according to the waiting time.
If there is any task that is dependent on another task, these tasks should be sorted based on their dependencies rather than their waiting times.

\textbf{(3) Provide a hospital queuing recommendation for the current patient.}

Finally, a hospital queuing recommendation with the sorted treatment tasks is performed for each patient by a mobile application interface.
Each patient can be invited to complete his treatment activities in the most convenient way with the least waiting time.
The detailed steps of the hospital queuing recommendation are presented in Algorithm \ref{alg32}.

\begin{algorithm}[!ht]
\caption{Process of the hospital queuing recommendation}
\label{alg32}
\begin{algorithmic}[1]
\REQUIRE ~\\
    $X$: the treatment data of the current patient;\\
    $PTTP_{RF}$: the trained PTTP model based on the RF algorithm.\\
\ENSURE ~\\
    $Ts(X)$: the recommended tasks with predicted waiting time.
\STATE create map $Ts(X) \leftarrow HashMap<string, double>$;
\FOR {each $Task_{i}$ in $X$}
    \STATE create array $U_{i}[] \leftarrow$ patients-in-waiting of $Task_{i}$;
    \FOR {each patient $U_{ik}$ in $U_{i}$}
    \STATE predict time consumption $T_{ik} \leftarrow PTTP_{RF}$;
    \ENDFOR
    \STATE calculate predicted waiting time $T_{i} \leftarrow \frac{1}{W_{i}} \sum_{k=1}^{m}{T_{ik}}$;
    \STATE append waiting time $Ts(X) \leftarrow <Task_{i}, T_{i}>$;
\ENDFOR
\STATE sort map $Ts(X)$ in an ascending order;
\FOR {each $<Task_{i}, T_{i}>$ in $Ts(X)$}
    \IF {($Task_{i}$ has dependent tasks)}
    \STATE put records of the dependent tasks before $Task_{i}$;
    \ENDIF
\ENDFOR
\RETURN $Ts(X)$.
\end{algorithmic}
\end{algorithm}

In Algorithm \ref{alg32}, $X$ contains the information of all of the treatment tasks for the current patient, such as task name, doctor name, and the patients waiting in the queue for each tasks.

\section{Parallel Implementation of the PTTP Algorithm and HQR System}
Massive historical treatment data (comprise more than 5 TB, and increase every day) are initially stored in HBase.
Then, the PTTP model and HQR system are parallelized in the Apache Spark cloud platform.
Thus, the performance of the algorithms is improved significantly.

\subsection{Parallel Implementation of the PTTP Model}
We parallelize the PTTP model on the Spark cloud platform.
A dual parallelization training process is performed.
The $k$ training subsets are trained in a parallel process, and $k$ CART regression trees are built at the same time.
Then, the $M$ variables in the training subsets are calculated in parallel in the node-splitting process of each tree.

The parallel training process of the PTTP model is implemented in the Spark computing cluster with the RDD programming model.
Distinct from the MapReduce model on the Hadoop platform, the intermediate results generated in the training process of the PTTP model are stored in the memory system on the Spark platform as RDD objects.

\begin{figure*}
\setlength{\abovecaptionskip}{0pt}
\setlength{\belowcaptionskip}{0pt}
\centering
\includegraphics[width=7.0in]{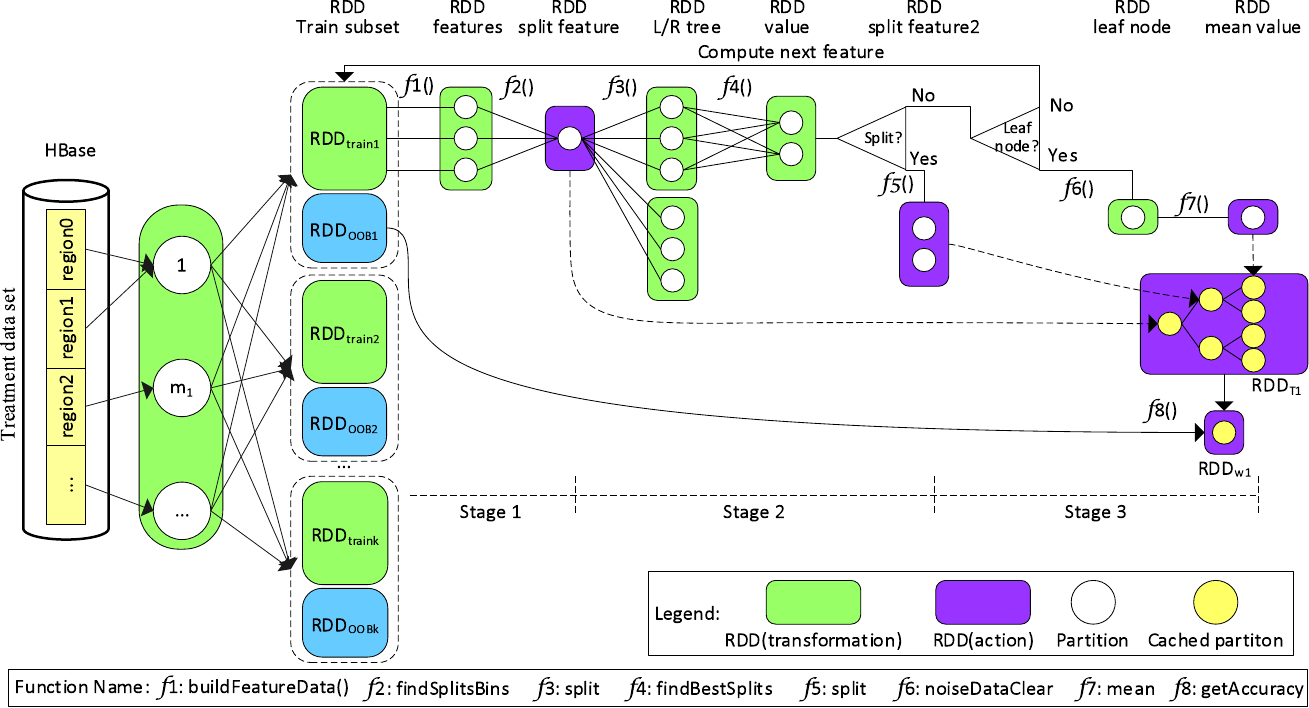}
\caption{Dual parallelization training process of the PTTP model}
\label{fig07}
\end{figure*}

The dual parallelization training process of the PTTP model is shown in Fig. \ref{fig07}.

Before the training process, the treatment data are loaded from HBase to the Spark Tachyon memory system as an RDD object.
An RDD object $RDD_{original}$ is defined to save the training dataset.
Then, $k$ training subsets are sampled as $k$ RDD objects from $RDD_{original}$; each of them is defined as $RDD_{traini}$.
Other $k$ RDD objects are created to save related OOB subsets; each of them is defined as $RDD_{OOBi}$.
The $k$ training subsets are allocated to $k$ map tasks at the same time and are allocated to multiple slave nodes.
Then, these training subsets are calculated in parallel with the RDD programming model including a series of operations.
Finally, $k$ regression tree models are obtained.

In the RDD programming model, each RDD object supports two types of operations, i.e., transformation and action.
Transformation operations include a series of operations on an RDD object, such as $map()$, $filter()$, $flatMap()$, $mapPartitions()$, $union()$, and $join()$.
Then, a new RDD object is returned from each transformation operation.
Action operations include a series of operations on an RDD object, such as $reduce()$, $collect()$, $count()$, $saveAsHadoopFile()$, and $countBykey()$, that compute a result and callback to the driver program or save it to an external storage system.
The detailed steps of the dual parallelization training process of the PTTP model are presented in Algorithm \ref{alg41}.

\begin{algorithm}[!ht]
\caption{Dual parallelization training process of the PTTP model}
\label{alg41}
\begin{algorithmic}[1]
\REQUIRE ~\\
    $RDD_{original}$: the treatment data loaded from HBase;\\
    $k$: the number of CART trees in the RF model.\\
\ENSURE ~\\
    $PTTP_{RF}$: the PTTP model based on the RF algorithm.
\STATE $trees \leftarrow$ SparkContext.parallelize($1$ to $k$, $slices$).\textbf{map}
\STATE \quad initialize subsets ($RDD_{traini}$, $RDD_{OOBi}$) $ \leftarrow$ \\
       \quad $randomSplit(RDD_{original})$;
\STATE \quad initialize feature subspaces $F_{sub} \leftarrow RDD_{traini}$;
\STATE \quad $F_{sub}$.parallelize($0$ to $F_{sub}.length$).\textbf{map}
\STATE \qquad calculate candidate split points $splits \leftarrow F_{sub}$;
\STATE \qquad calculate best split $Split_{best} \leftarrow$ $node.findBest$-\\
       \qquad $Splits(F_{sub}, splits).sortByKey().top(1)$;
\STATE \qquad append $node$ to $tree_{i}$;
\STATE \qquad split data to two forks ($RDD_{L}$, $RDD_{R}$) $\leftarrow$ $node$. \\
       \qquad split($Split_{best}$);
\STATE \qquad \textbf{if}{($\Phi(RDD_{L}|RDD_{R}) \geq \Phi(Split_{best})$)} \textbf{then}
\STATE \qquad \quad append $subnode$ to $node$ as multi-branch;
\STATE \qquad \quad split data to two forks ($RDD_{L2}$, $RDD_{R2}$) $\leftarrow$ \\
       \qquad \quad $subnode.split(Split_{best2})$;
\STATE \qquad \textbf{else}
\STATE \qquad \quad clean noisy data $RDD_{leaf}$ $\leftarrow$ $RDD_{L}|RDD_{R}.$ \\
       \qquad \quad $noisyDClean()$;
\STATE \qquad \quad calculate mean value of leaf node $c$ $\leftarrow$ $RDD_{leaf}.$ \\
       \qquad \quad $mean()$;
\STATE \qquad \textbf{endif}
\STATE \quad \textbf{endmap}.groupBykey().reduce();
\STATE \quad build CART $tree_{i} \leftarrow$ new $CARTModel(nodes)$;
\STATE \quad calculate accuracy $CA_{i}$ $\leftarrow$ $tree_{i}.getAc(RDD_{OOBi})$;
\STATE \quad return ($tree_{i}$, $CA_{i}$);
\STATE \textbf{endmap}.collect();
\STATE $PTTP_{RF} \leftarrow$ RandomForestModel($trees$);
\RETURN $PTTP_{RF}$.
\end{algorithmic}
\end{algorithm}

The training processes of each training subset $RDD_{traini}$ and the OOB subset $RDD_{OOBi}$ comprise the following stages.

In stage 1, there are $buildFeatureData()$ and $findSplitsFeature()$ functions, which perform a transformation operation and an action operation, respectively.
In the $buildFeatureData()$ function, feature subspaces of $RDD_{traini}$ are mapped to a new RDD object with $M$ partitions, which refer to the $M$ feature variables.
The loss function of each feature variable subspace and each potential split point value of the variable are calculated.
In the $findSplitsFeature()$ function, the results of the variable's loss function are sorted, and the feature variable with the least value is selected as the first node of CART tree $T_{i}$, which is created as RDD object $RDD_{Ti}$.

In stage 2, there are two $split()$ functions and a $findBestSplits()$ function.
In the first $split()$ function, the training subset $RDD_{traini}$ is split into two forks by a split point in the current feature subspace, which is shown as $RDD_{L/Rtree}$ in Fig. \ref{fig08}.
For each branch, there is a $findBestSplits()$ function.
In the $findBestSplits()$ function, the same feature variables continue to be selected, and the results of sets of the potential splitting values for the current feature subspace are calculated.
The best split point is obtained for the data in the branch, such as $RDD_{splitfeature2}$.
Then, if the split value is greater than the father node, the branch continues to split by the current feature variable and the best split point in the second $split()$ function.
Otherwise, the other remaining feature variables continue to be computed.
If the current tree node is not a leaf node, repeat stages (1 - 2) to compute the next feature, except for the features that exist in tree nodes.
Alternatively, if the current node is a leaf node, go to stage 3.

In stage 3, there are a $noisyDataClear()$ function and a $mean()$ function.
The noisy data of each leaf node are cleaned in the $noisyDataClear()$ function.
Then, in the $mean()$ function, the average value of the data is calculated, which is the value of the leaf node of the $RDD_{Ti}$.

The splitting process is repeated until all of the feature variables are calculated.
A tree model $RDD_{Ti}$ for the training subset $RDD_{traini}$ is trained.
Finally, the OOB subset $RDD_{OOBi}$ against the training subset $RDD_{traini}$ is used to test the accuracy of the tree $RDD_{Ti}$, and the accuracy of $RDD_{Ti}$ is computed as the weight in a $getAccuracy()$ function.
Taking advantage of the cloud-computing platform and a distributed memory management mechanism, the performance of the parallel method is improved evidently.

\subsection{Parallel Implementation of the HQR System}
\begin{figure}[!ht]
\setlength{\abovecaptionskip}{0pt}
\setlength{\belowcaptionskip}{0pt}
\centering
\includegraphics[ width=3.5in]{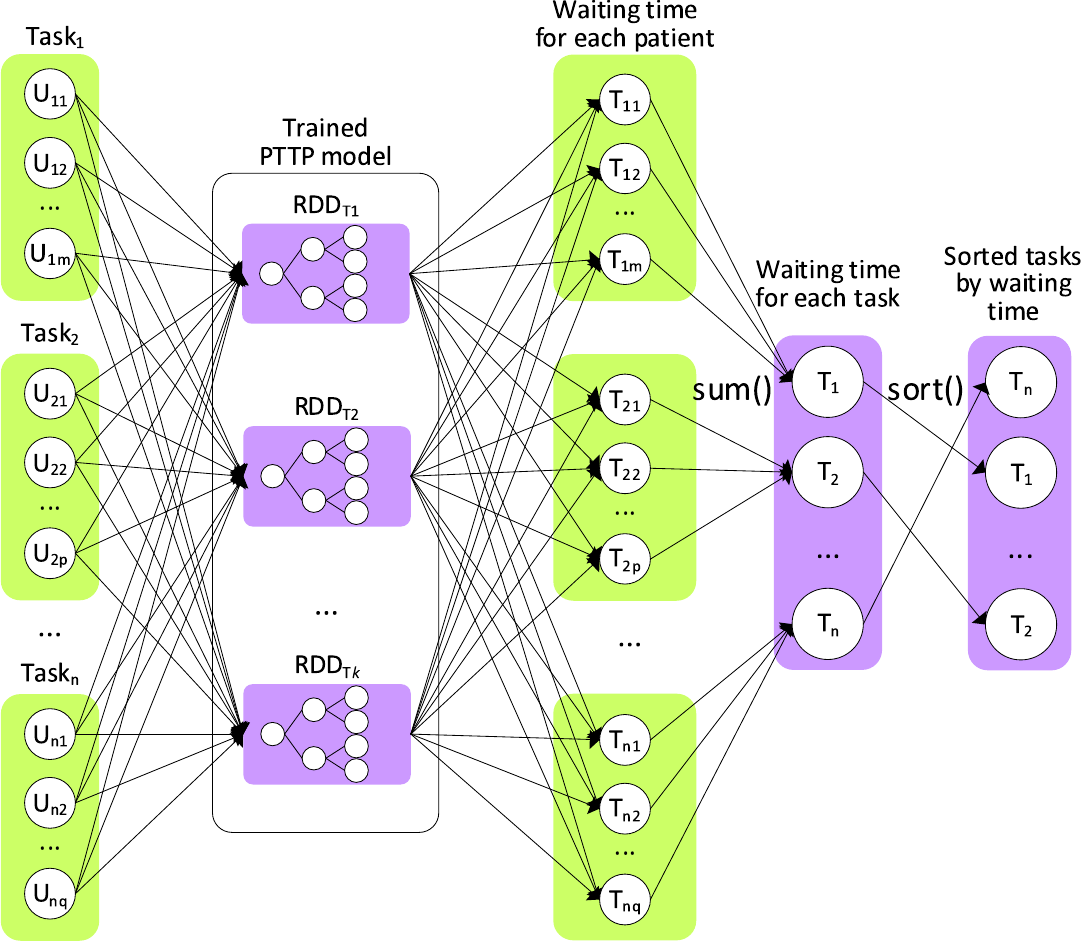}
\caption{Parallelization recommendation process of the HQR system}
\label{fig08}
\end{figure}

Usually, there are a number of treatment tasks for each patient, and many patients waiting in the queue of each treatment task.
Therefore, a parallel HQR system is implemented for each patient if there is more than one treatment task for the patients.
The process of the parallel HQR system is shown in Fig. \ref{fig08}.

Assume that there are $n$ treatment tasks for the current patient to complete and that there is a number of patients waiting in the queue of each treatment task.
In the parallelization solution, $n$ RDD objects are created to refer to the $n$ treatment tasks.
There is a number of partitions in each RDD object that refer to patients waiting in the queue of each task.
Let partition $U_{ij}$ be the $j$th patient waiting for the $i$th treatment task.

Step 1: For each patient $U_{ij}$ in a task $Task_{i}$, the time consumption of the patient might generate in the $i$th task, as predicted by the trained PTTP model.
In this step, the time consumption for each patient $U_{ij}$ is calculated with the $k$ trained CART trees of the RF-based PTTP model in a $shuffle()$ function, and the predicted patient treatment time consumption $T_{ij}$ is derived.

Step 2: The patient treatment time consumption of all of the patients in each task is added in a $sum()$ function, and the predicted waiting time $T_{i}$ of each task is obtained.
An RDD object $(Task_{i}, T_{i})$ is created for each task.

Step 3: The predicted waiting times for all of the tasks for the current patient are sorted in ascending order with a $sort()$ function.
A new RDD object $Ts$ is created to save the sorted waiting times of all of the treatment tasks.
Hence, the parallel hospital queuing recommendation schema for the current patient is performed.
The detailed steps of the parallel HQR algorithm are presented in Algorithm \ref{alg42}.

\begin{algorithm}[!ht]
\caption{Parallelization recommendation process of the HQR algorithm}
\label{alg42}
\begin{algorithmic}[1]
\REQUIRE ~\\
    $RDD_{Tasks}$: the treatment tasks data of the current patient;\\
    $PTTP_{RF}$: the trained PTTP model based on the RF algorithm.
\ENSURE ~\\
    $Ts$: the recommended treatment tasks list with predicted waiting time.
\STATE $Ts \leftarrow RDD_{Tasks}$.\textbf{map}
\STATE \quad $Task_{i} \Rightarrow$
\STATE \quad $U_{i}$ $\leftarrow getWaitingPatients(Task_{i})$;
\STATE \quad $T_{i} \leftarrow U_{i}$.\textbf{map}
\STATE \qquad $U_{ik} \Rightarrow$
\STATE \qquad predict time consumption $T_{ik} \leftarrow PTTP_{RF}.predict(U_{k}.vars)$;
\STATE \quad \textbf{endmap}
\STATE \quad ($U_{ik}, T_{ik}$).collect().reduce();
\STATE \quad return predicted waiting time $(Task_{i}, T_{i})$;
\STATE \textbf{endmap}
\STATE ($Task_{i}, T_{i}$).reduceByKey();
\STATE sort tasks list $Ts \leftarrow Ts.sortByTime()$;
\FOR {each $(Task_{i}, T_{i})$ in $Ts(X)$}
    \IF {($Task_{i}$ has dependent tasks)}
    \STATE put records of the dependent tasks before $Task_{i}$;
    \ENDIF
\ENDFOR
\RETURN $Ts$.
\end{algorithmic}
\end{algorithm}

\section{Experiments and Applications}
In this section, the accuracy and performance of the proposed algorithm are evaluated through a series of experiments.
The algorithm is applied to an actual hospital project in China.
Section 5.1 presents the experimental settings.
The experiment result analysis of the PTTP algorithm and the HQR system are presented in Section 5.2, Section 5.3 presents the accuracy and robustness evaluation, and performance evaluation is provided in Section 5.4.

\subsection{Experiment and Application Setup}
The HQR system consists of two main modules: a decision maker and recommendation module and a mobile application interface module.
In the decision maker and recommendation module, treatment data are transmitted to the HBase database in NSCC from hospitals regularly.

The system and experiments are performed on a Spark cloud platform, which is constructed at the National Supercomputing Center in Changsha to achieve the aforementioned goals.
Each computing node runs Linux operating system Ubuntu 12.04.4, with 2 Intel Xeon Westmere EP CPUs, 6 cores, 2.93GHZ, and 48GB memory.
All of the nodes are connected by a high-speed Gigabit network and are configured with Hadoop 2.6.0 and Spark 1.6.0. The algorithm is implemented in Java 1.7.0 and Scala 2.11.7.
In our experiments, datasets covering three years (2012 - 2014) are chosen from an actual hospital application, as shown in Table \ref{table51}.

\begin{table}[!h]
\setlength{\abovecaptionskip}{0pt}
\setlength{\belowcaptionskip}{0pt}
\tabcolsep1pt
\renewcommand{\arraystretch}{1.3}
\caption{Datasets from an actual hospital application}
\label{table51}
\begin{tabular}{p{0.5in}  p{0.7in}  p{0.7in}  p{0.7in}  p{0.7in}c}
\hline
Years & Departments & Tasks & Instances & Data Size \\
\hline
2012 & 285 & 14,481 & 189,186,143 & 1.4 TB \\
2013 & 299 & 14,769 & 229,873,259 & 1.6 TB \\
2014 & 294 & 15,012 & 238,935,397 & 2.0 TB \\
\hline
\end{tabular}
\end{table}

In Table \ref{table51}, the departments of the hospital include the financial room, the Emergency Department (ED), CT scan, MR scan, B-model ultrasound, color Doppler ultrasound, nuclear medicine, and the pharmacy.
There are various treatment tasks in each department.

\subsection{Experiment Result Analysis}
We analyze the patient treatment time consumption of the CT scan task with time factors and patient characteristics.
Because of the content of the activities and various circumstances, the patient treatment time consumption of treatment tasks in each department can vary.
At the same time, the time consumption in the same department might be different due to the different treatment tasks, different periods, and different conditions of patients.

\subsubsection{Treatment Time Consumption with Time Factors}

\begin{figure}[!ht]
\setlength{\abovecaptionskip}{0pt}
\setlength{\belowcaptionskip}{0pt}
\centering
\includegraphics[width=3.5in]{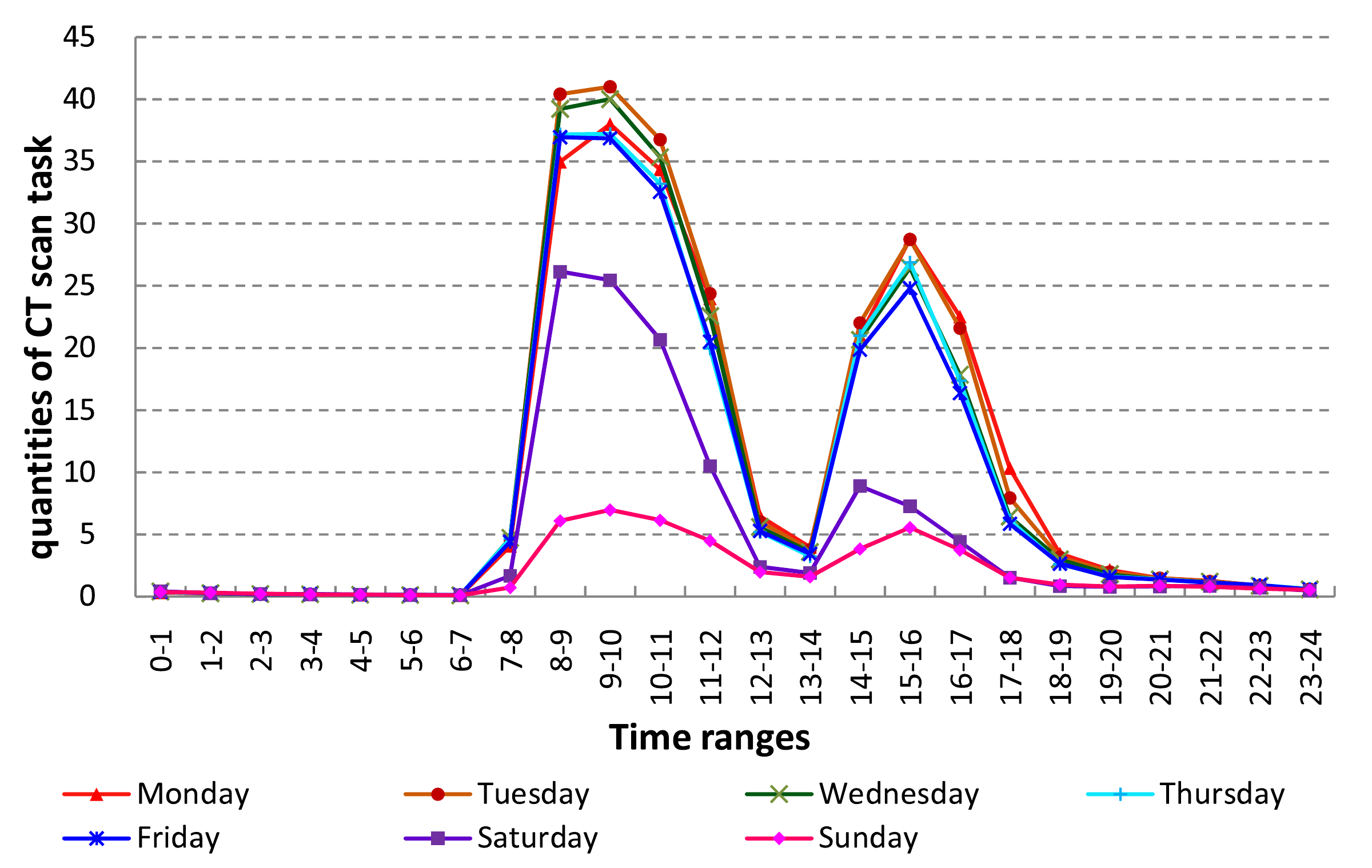}
\caption{CT scan task quantities in each week period}
\label{chart01}
\end{figure}

The CT scan treatment task quantities are depicted in Fig. \ref{chart01}.
As seen in Fig. \ref{chart01}, there are two peaks of the CT scan task every day.
The first peak comes from 8 am to 11 am, and the second peak comes from 2 pm  to 5 pm.
The nadir point of each day is in the range of 0 am -7 am in the morning, 12 pm to 1 pm at noon, and 6 pm to 11 pm in the evening.
The overall number of patients per weekend day is less than that on individual weekdays.

\begin{figure}[!ht]
\setlength{\abovecaptionskip}{0pt}
\setlength{\belowcaptionskip}{0pt}
\centering
\includegraphics[width=3.5in]{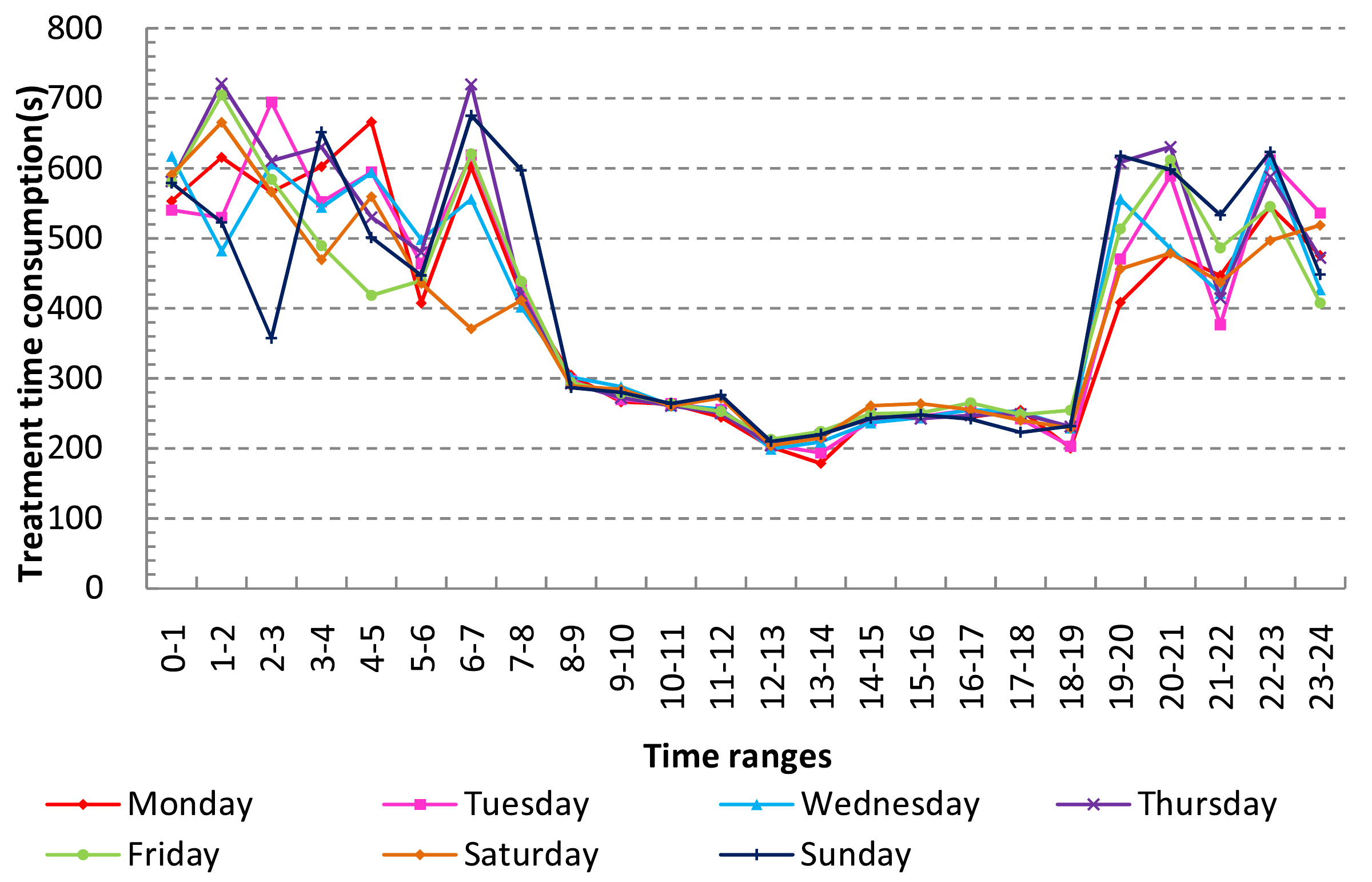}
\caption{Treatment time consumption of the CT scan task with time factors (part)}
\label{chart02}
\end{figure}

After the training process of the PTTP algorithm, the time consumptions of all of the treatment tasks in the experiment are trained. The time consumption of CT scan task with time factors (part) is shown in Fig. \ref{chart02}.

Each point in Fig. \ref{chart02} refers to a value of one leaf node in the regression trees of the PTTP model.
Consider 9 am on a weekday for a CT scan task to be an example of a peak time scenario; the average output of a CT scan operation is approximately 40 every day.
There are 43,200 records at the leaf nodes of the CART tree model.
The time consumption is close to 240 s (approximately 4.0 min) for a CT scan task.
Conversely, at the nadir point, there are 0 or 1 CT scan tasks in each hour.
There are 0 - 1095 (1 $\times$ 365 days $\times$ 3 years) records at the leaf node of the tree model.

Obviously, because there are approximately 43,200 (40 $\times$ 365 days $\times$ 3 years) records at the leaf node for peak time case, the value of trained treatment time consumption is smooth and steady.
At the nadir point, the value of trained treatment time consumption is undulate because of the small number of training samples.
Consequently, having fewer records in each leaf node of the tree model results in less accuracy.

\subsubsection{Treatment Time Consumption with Patient Characteristics}

\begin{figure}[!ht]
\setlength{\abovecaptionskip}{0pt}
\setlength{\belowcaptionskip}{0pt}
\centering
\includegraphics[width=3.5in]{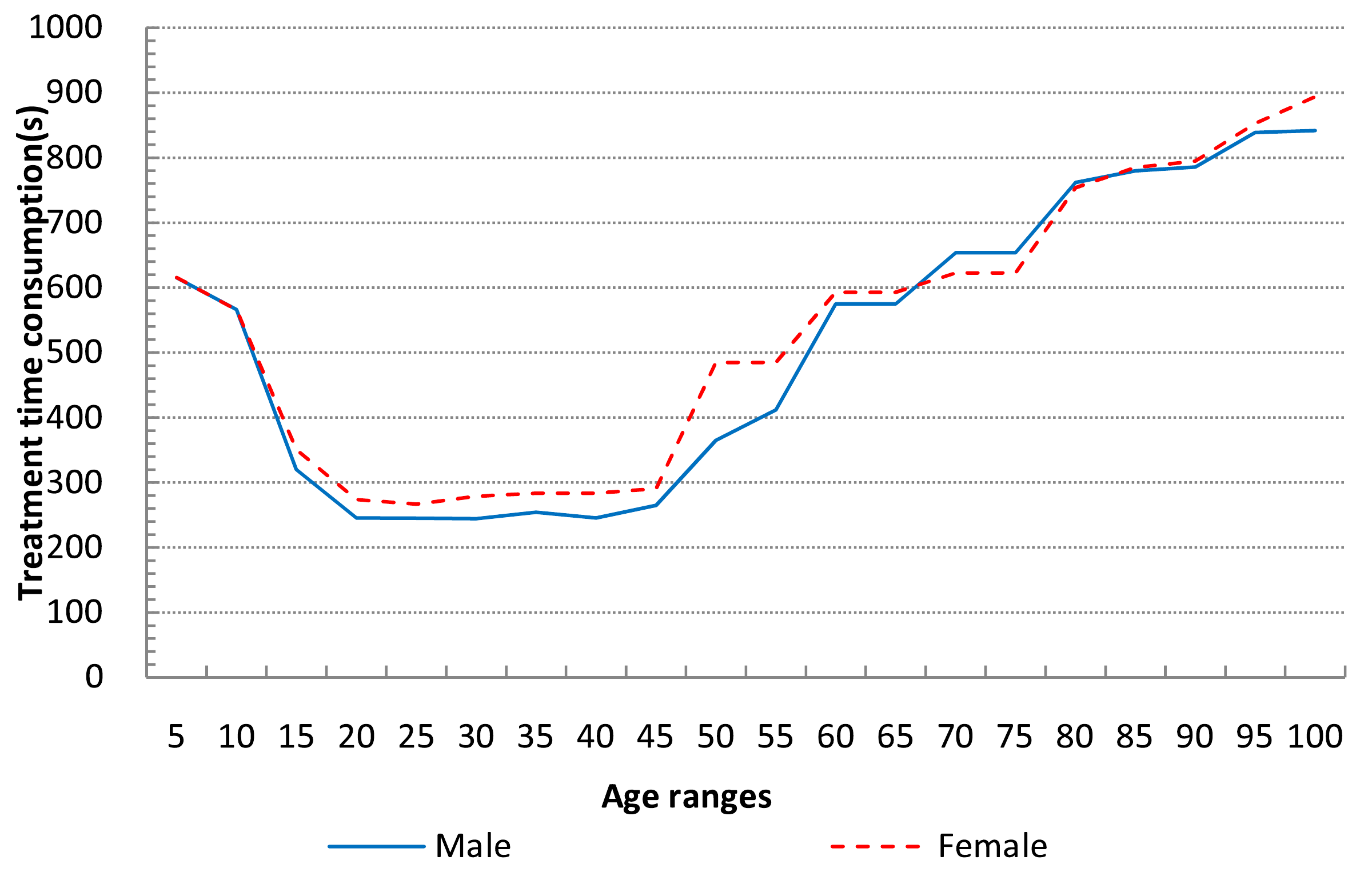}
\caption{Treatment time consumption of the CT scan task with patient characteristics (part)}
\label{chart03}
\end{figure}

The treatment time consumption of a CT scan task with patient characteristics (part) is shown in Fig. \ref{chart03}.

As seen in Fig. \ref{chart03}, for patients with ages ranging from 20 to 40, time consumption of each CT scan task is approximately 245 s (approximately 4.1 min) for both men and women.
As age increases, the time required for each patient's CT scan task increases.
For example, the time consumption for a male patient at age 90 is approximately 786 s (approximately  13.1 min).
At the same time, generally speaking, the time consumption for a female patient is greater than that for a male in the same age range.

\subsubsection{HQR system in a Mobile Application}
\begin{figure}[!ht]
\setlength{\abovecaptionskip}{0pt}
\setlength{\belowcaptionskip}{0pt}
\centering
 \subfigure[Recommended tasks list]{
 \label{app1:a}
 \includegraphics[width=1.6in]{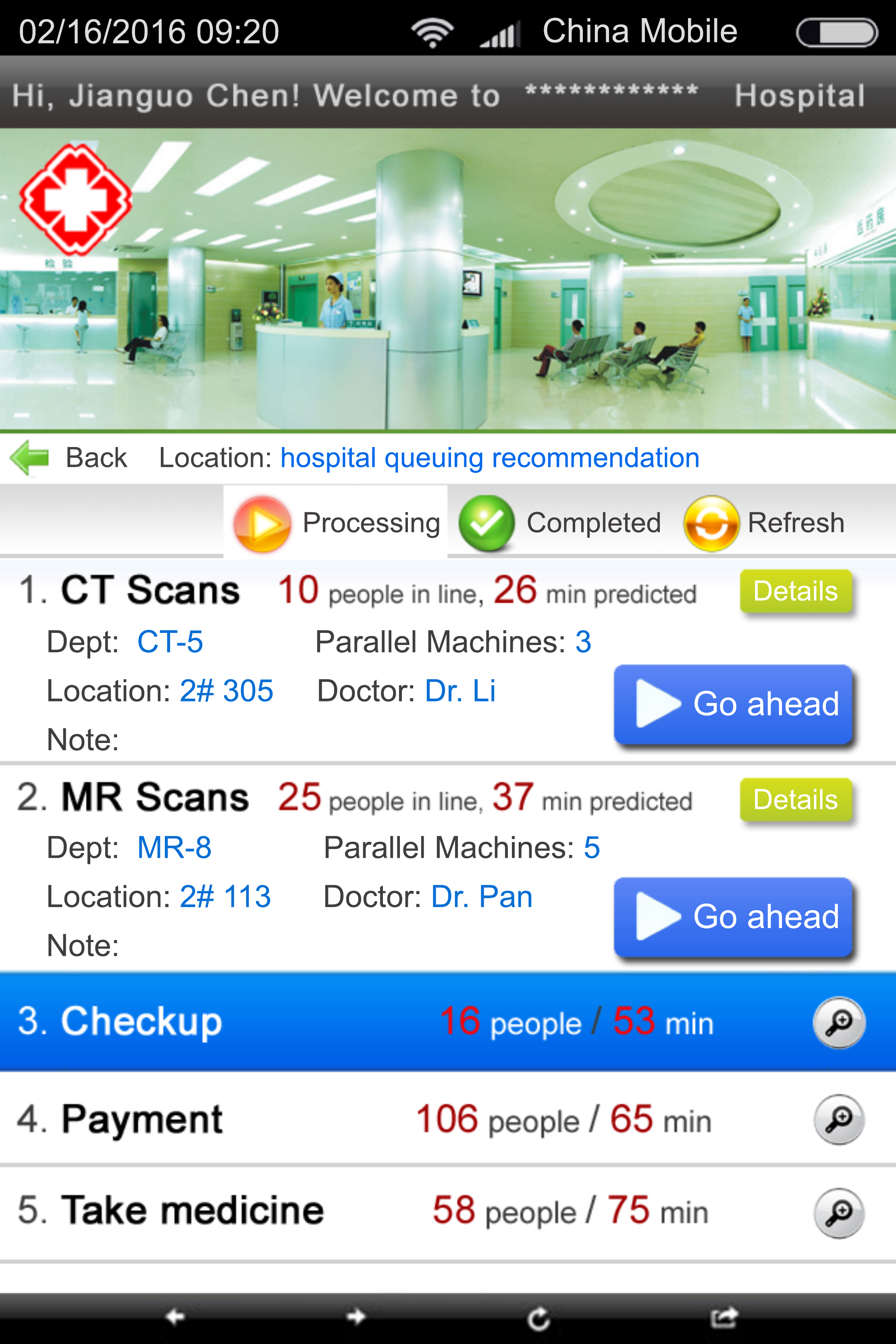}}
 \hspace{0.1in}
 \subfigure[Details of the waiting queue]{
 \label{app1:b}
 \includegraphics[width=1.6in]{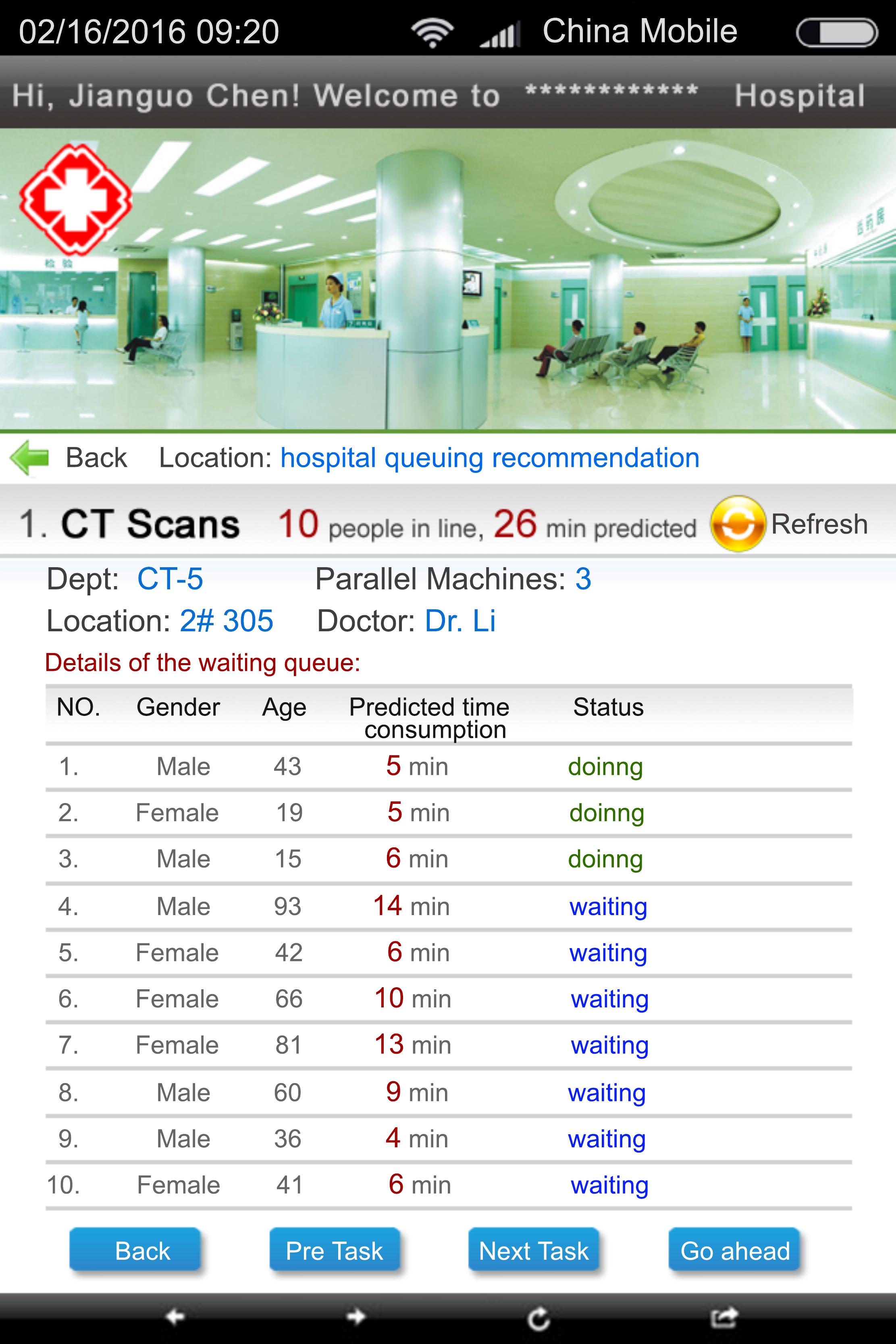}}
 \caption{Mobile interfaces of the HQR system}
 \label{app1}
\end{figure}

To elaborate the working of the HQR system, an example experiment is discussed below.
One patient is considered to be an example scenario. The patient must undergo various treatment tasks, such as a doctor checkup, a CT scan, an MR scan, a pharmacy visit to obtain prescribed medicines, and a payment task.
As mentioned above, a set of treatment tasks for the current patient is submitted to the decision maker and recommendation module through a mobile interface. The mobile interface of the HQR system is shown in Fig. \ref{app1}.
Because the language of the mobile application is Chinese, we have translated the language from Chinese to English.

The predicted waiting time of all of the treatment tasks is calculated by the PTTP model. Then, a treatment recommendation with the least waiting time is advised.
Fig. \ref{app1:a} shows that there are 10 people waiting for the CT scan before the current patient (including the people waiting in the queue and in processing), and the predicted waiting time is 26.0 min.
Fig. \ref{app1:b} shows the details of the waiting queue for the CT scan.
We can see the characteristics, predicted time consumption, and the status of each person in the queue.
For example, the treatment time consumption of a 15-year-old male is 6.0 min, which is close to the trained time consumption of 350 s (shown in Fig. \ref{chart03}).
The total predicted time consumption of 10 people is 78.0 min, and there are 3 machines available in parallel.
Therefore, the predicted waiting time of the current patient is 26.0 min.
Moreover, the status of the waiting queue is updated in real-time.
The experimental results show that the HQR system provides a recommendation with an effective treatment plan for patients to minimize their wait times in hospitals.

\subsubsection{Average Waiting Time for Patients}
\begin{figure}[!ht]
\setlength{\abovecaptionskip}{0pt}
\setlength{\belowcaptionskip}{0pt}
\centering
\includegraphics[width=3.0in]{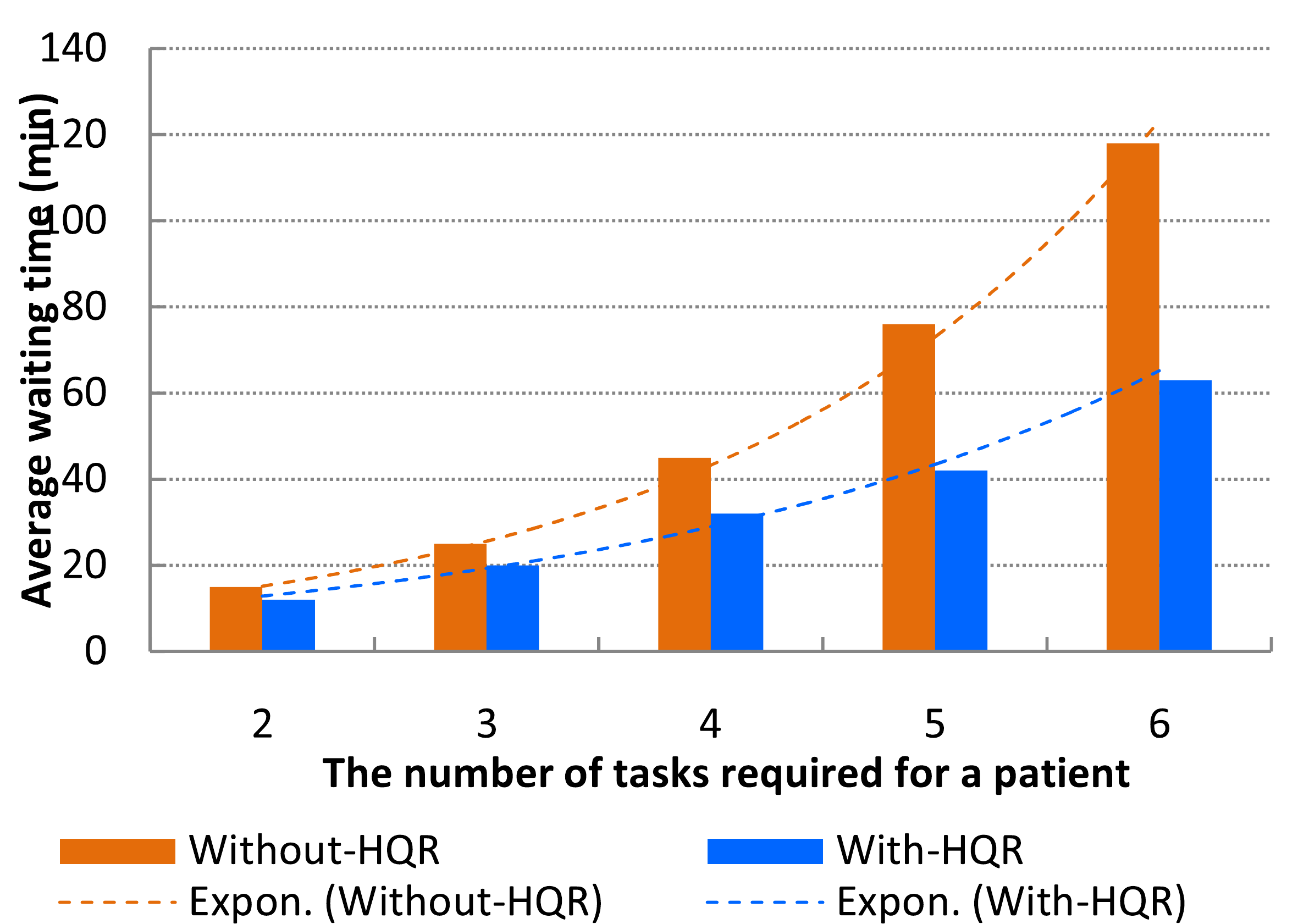}
\caption{Average waiting time for patients}
\label{chart04}
\end{figure}

To evaluate the efficiency of our HQR system, various experiments about average waiting time for patients in the with-HQR case with that in the without-HQR case are performed.
Each case is under the treatment data with 5000 patients and 20,000 treatment records.
We accounted and compared the average waiting time of patients in the with-HQR case with that in the without-HQR case.
The results of comparison are presented in Fig. \ref{chart04}.

It is easy to observe from Fig. \ref{chart04} that the advantage of the average waiting time of patients in cases of with-HQR is greater than in cases of without-HQR.
Moreover, the more patients treatment tasks are, the more obvious is for this advantage.
When the number of tasks required for each patient is equal to 2, the average waiting time of each patient is approximately 15 min in the without-HQR case (the original case), while 12 min in the with-HQR case.
When there are 6 treatment tasks required for each patient, the average waiting time is approximately 118 min in the former case, while 63 min in the latter case.

\subsection{Accuracy and Robustness Analysis}
To evaluate the accuracy and robustness of our improved-RF-based PTTP algorithm, we implemented the PTTP algorithm based on the original random forest (refereed as PTTP-ORF).
The accuracies of the PTTP algorithm and PTTP-ORF algorithm are analyzed under different ratios of noisy data.

\subsubsection {Results Evaluation of Noise Removal}
In Section 3.2.2, a noise removal method is introduced in the training process of the regression tree model.
The effect of noise removal is validated and analyzed.
Six groups of leaf node data in the regression tree models are discussed in experiments, the specific conditions of the six groups of leaf nodes are shown in Table \ref{table54}.

\begin{table}[!ht]
\setlength{\abovecaptionskip}{0pt}
\setlength{\belowcaptionskip}{0pt}
\tabcolsep1pt
\renewcommand{\arraystretch}{1.3}
\caption{Specific conditions of six leaf nodes in the experiments}
\label{table54}
\begin{tabular*}{3.5in}{p{0.7in} p{2.8in}}
\hline
Leaf Node & Condition of the leaf node\\
\hline
CT-1   & \{Task: CT scan, Gender: Male, Age range: 25-45\}.\\
CT-2   & \{Task: CT scan, Gender: Male, Age range: 65-85, Week: Monday, Time range: 8-12\}.\\
MR-1   & \{Task: MR, Gender: Male, Age range: 20-45\}.\\
MR-2   & \{Task: MR, Gender: Male, Age range: 65-85, Week: Monday-Friday, Time range: 8-12\}.\\
\hline
\end{tabular*}
\end{table}

\begin{figure}[!ht]
\setlength{\abovecaptionskip}{0pt}
\setlength{\belowcaptionskip}{0pt}
\centering
\includegraphics[width=3.0in]{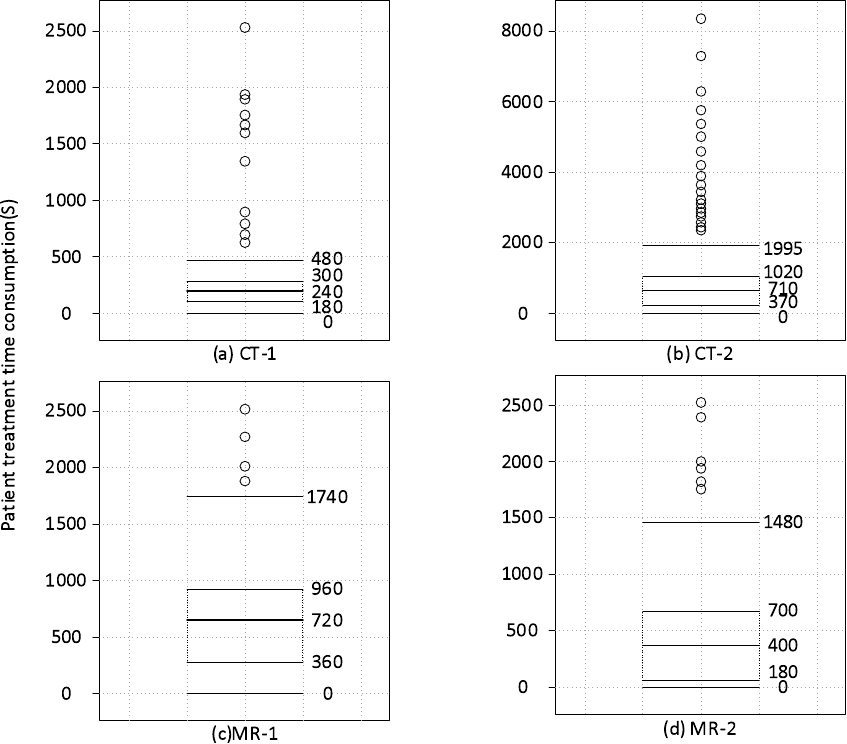}
\caption{Noisy data removal results for the PTTP algorithm}
\label{chart05}
\end{figure}

The results of noise removal for the PTTP algorithm are presented in Fig. \ref{chart05}.

Fig. \ref{chart05}(a) is a box plot of a leaf node with the condition of ``CT-1".
The patient treatment time consumption in this case is between 0 and 2500 s (approximately 0.0 - 41.6 min).
The boundaries of the box plot in this case are 0 and 480 s (approximately 8.0 min), and the median value is 240 s (approximately 4.0 min).
That is, most of the patient treatment time consumption data are in this range, which is understandable for people in the 25-45 age range in the treatment operation of a CT scan task.
In Fig. \ref{chart05}(b), time consumption is in the range 0 - 8000 s (approximately 0.0 - 133.3 min) for male aged 65 - 85 in the CT scan task.
After noise removal, the time range is changed to 0 - 1995; the median value is 710 s (approximately 11.8 min).
In Fig. \ref{chart05}(c), the time consumption range is 0 - 1740 s (approximately 0.0 - 29.0 min) after noise removal, rather than the range of 0 - 2500 s.
The median value is 720 s (approximately 12.0 min) for one treatment of the MR scan task.

\begin{figure}[!ht]
\setlength{\abovecaptionskip}{0pt}
\setlength{\belowcaptionskip}{0pt}
\centering
\includegraphics[width=3.0in]{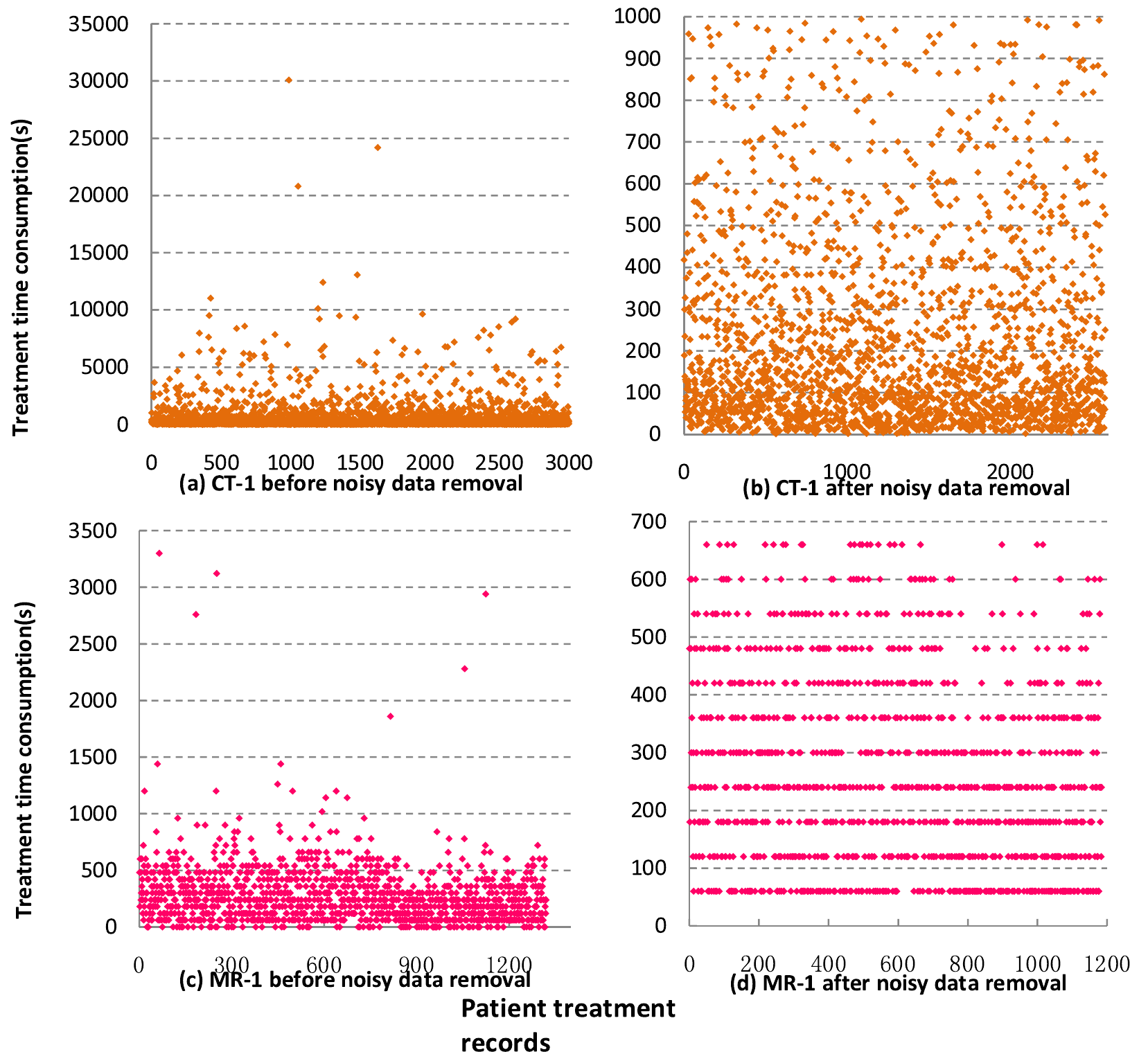}
\caption{Examples of noisy data removal from patient treatment time consumption}
\label{chart06}
\end{figure}

Two examples of noisy data removal from patient treatment time consumption are shown in Fig. \ref{chart06}.

Fig. \ref{chart06}(a) and Fig. \ref{chart06}(b) show the patient treatment time consumption of a leaf node before and after noise removal.
After noise removal, the range of the value is changed from (0 - 35,000) to (0 - 1000), and the value range decreases by 97.14\%.
The number of records decreases from 3000 to 2582.
Namely, the number of noisy data points is equal to 418, and the noise rate is 13.93\%.
Fig. \ref{chart06}(c) and Fig. \ref{chart06}(d) depict the patient treatment time consumption of another leaf node before and after noise removal.
After noise removal, the range of the value is changed from (0 - 3500) to (0 - 700), and the value range decreases by 80.00\%.
The number of records decreases from 1320 to 1185.
The number of noisy data points is equal to 135, and the noise rate is 10.23\%.
Summarizing, after noise removal, the value ranges of patient treatment time consumption obviously decrease.

\subsubsection{Algorithm Accuracy Analysis with Different Tree Scales}
\begin{figure}[!ht]
\setlength{\abovecaptionskip}{0pt}
\setlength{\belowcaptionskip}{0pt}
\centering
\includegraphics[height = 2.25in, width=3.5in]{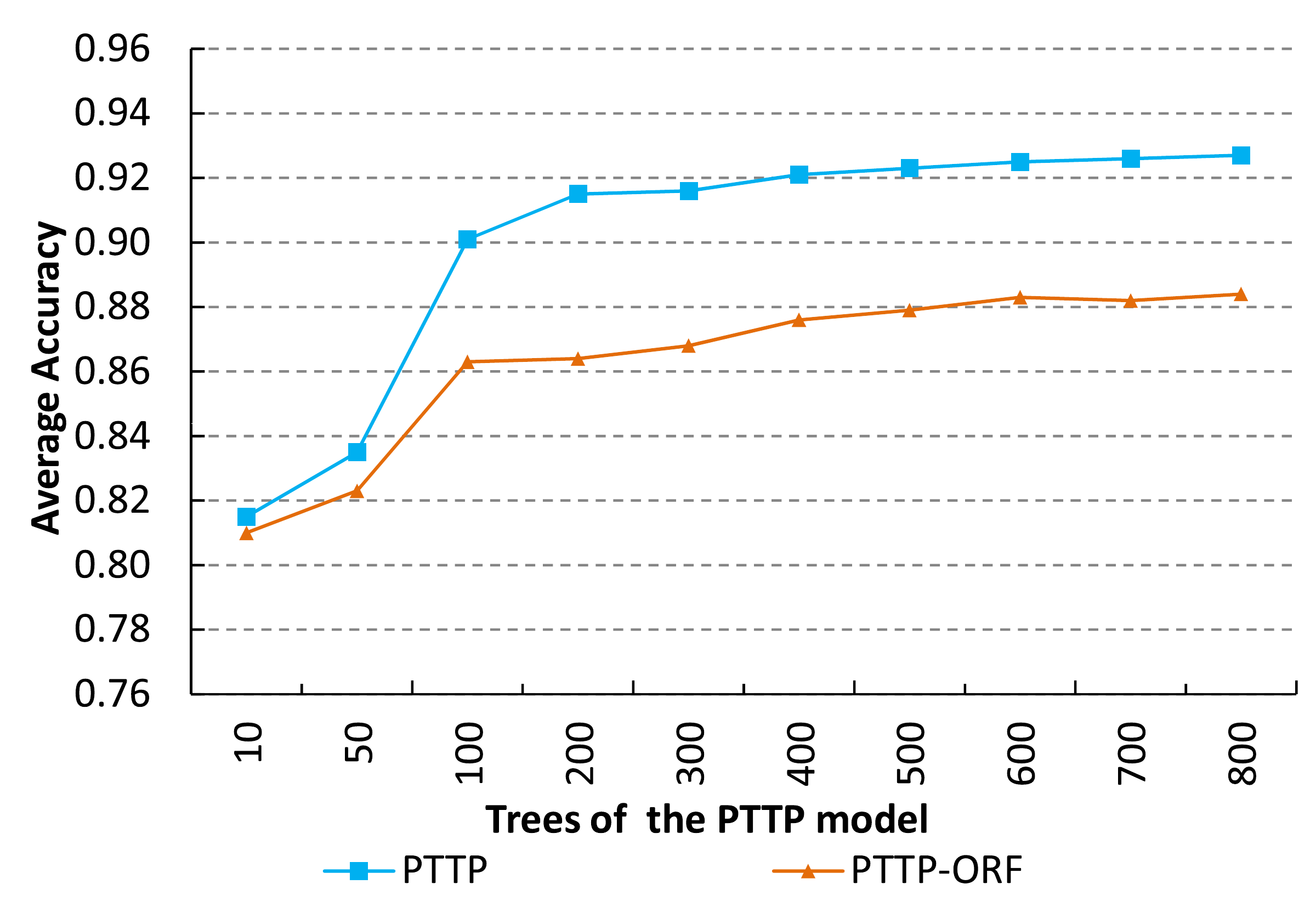}
\caption{Accuracy of different algorithms with different tree scales}
\label{chart07}
\end{figure}

To illustrate the accuracy of the PTTP algorithm, various experiments are performed on the dataset shown in Table \ref{table51}.
Each case is under different scales of the decision tree.
By counting the average accuracies of the algorithms, the different accuracies of various environments are compared and analyzed.
The results are presented in Fig. \ref{chart07}.

Fig. \ref{chart07} shows that the average accuracy of the PTTP algorithm based on different improved random forest algorithms is not high when the number of regression trees in each algorithm is equal to 10.
With an increase in the number of decision trees, the average accuracy increases gradually and tends toward a convergence condition.
The accuracy of the PTTP algorithm is greater than that of PTTP-ORF by 3.72\% on average and 5.10\% in the best case, when the number of decision trees is equal to 200.
Consequently, compared with PTTP-ORF, the PTTP algorithm, which has been optimized in four aspects, can significantly increase the accuracy.

\subsubsection{Algorithm Accuracy Analysis under Different Noise Ratios}
\begin{figure}[!ht]
\setlength{\abovecaptionskip}{0pt}
\setlength{\belowcaptionskip}{0pt}
\centering
\includegraphics[width=3.5in]{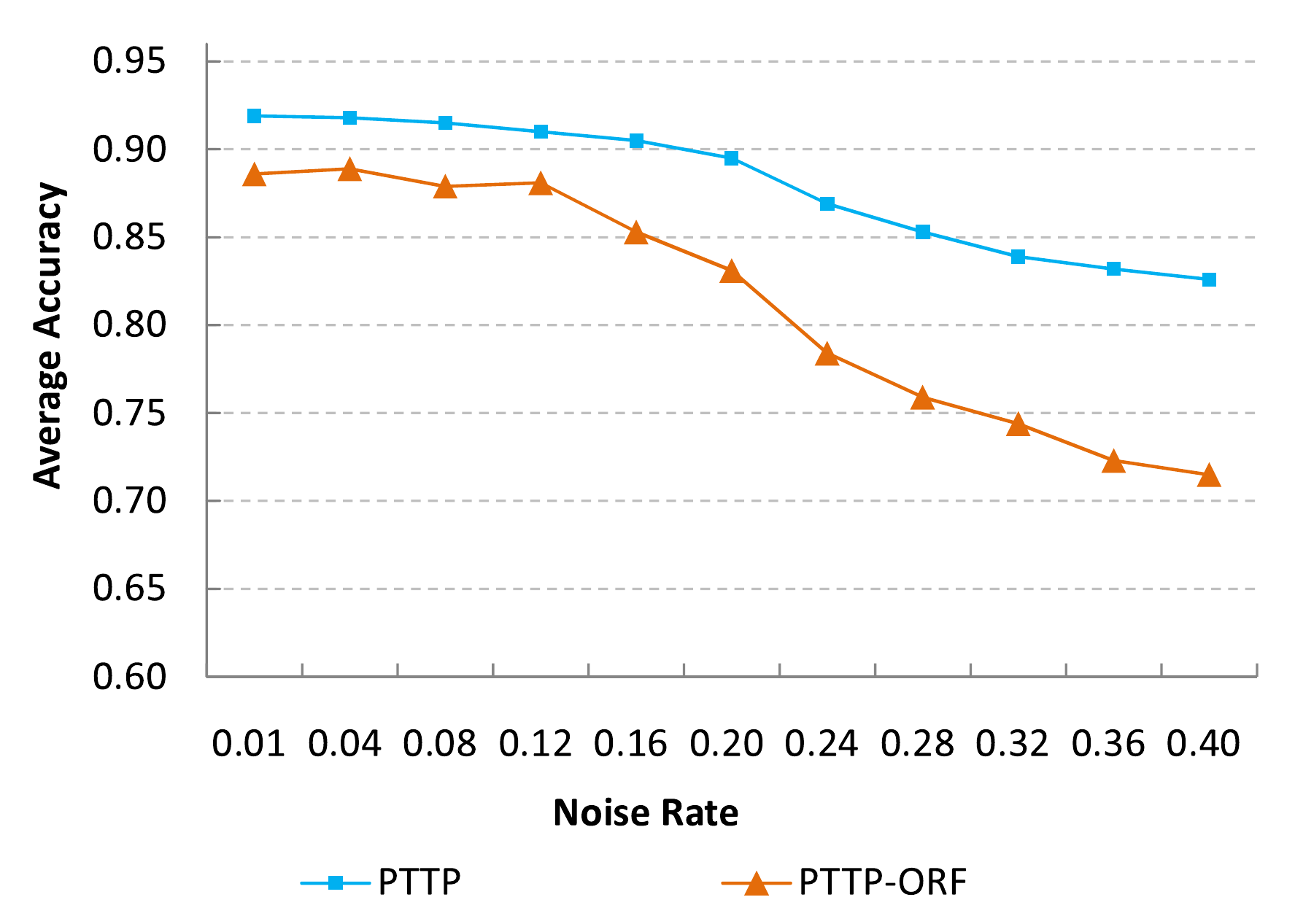}
\caption{Accuracy of different algorithms under different noise ratios}
\label{chart08}
\end{figure}

To demonstrate the accuracy of our algorithm, we conduct experiments with algorithms, such as the PTTP and PTTP-ORF.
We construct the noisy data by modifying the values of the original data randomly according to different noise ratio requirements.
The scales of the noise ratios are located in the range of \{1\%, 4\%, 8\%, 12\%, 16\%, 20\%, 24\%, 28\%, 32\%, 36\%, 40\%\}.
The number of training samples in the cases is 100,000, and the number of regression trees in the random forest model is 500.
The result of comparative analysis is presented in Fig. ~\ref{chart08}.

Fig. ~\ref{chart08} states that in each case, when the proportion of noisy data increases, the average accuracy of PTTP-ORF decreases quickly.
When the scale of noisy data increases from 1\% to 40\%, the accuracy of PTTP-ORF decreases from 88.70\% to 74.50\%.
Therefore, noisy data have a significant degree of influence on PTTP-ORF.
Accuracy of PTTP-ORF is influenced by a large volume of noisy data.
In addition, as the proportion of noisy data increases, the tendency of the accuracy of our PTTP algorithm decrease is steady.
When the proportion of noisy data increases from 1\% to 50\%, the average accuracy of PTTP decreases from 91.90\% to 82.60\%.

Obviously, the average accuracy of PTTP is greater than that of the other two algorithms under each condition of noise ratio.
Consequently, the PTTP algorithm can reduce the influence of noisy data effectively and achieve good robustness.

\subsection{Performance Evaluation}
\subsubsection{Performance Evaluation of the PTTP Algorithm}
\begin{figure}[!ht]
\setlength{\abovecaptionskip}{0pt}
\setlength{\belowcaptionskip}{0pt}
\centering
\includegraphics[width=3.5in]{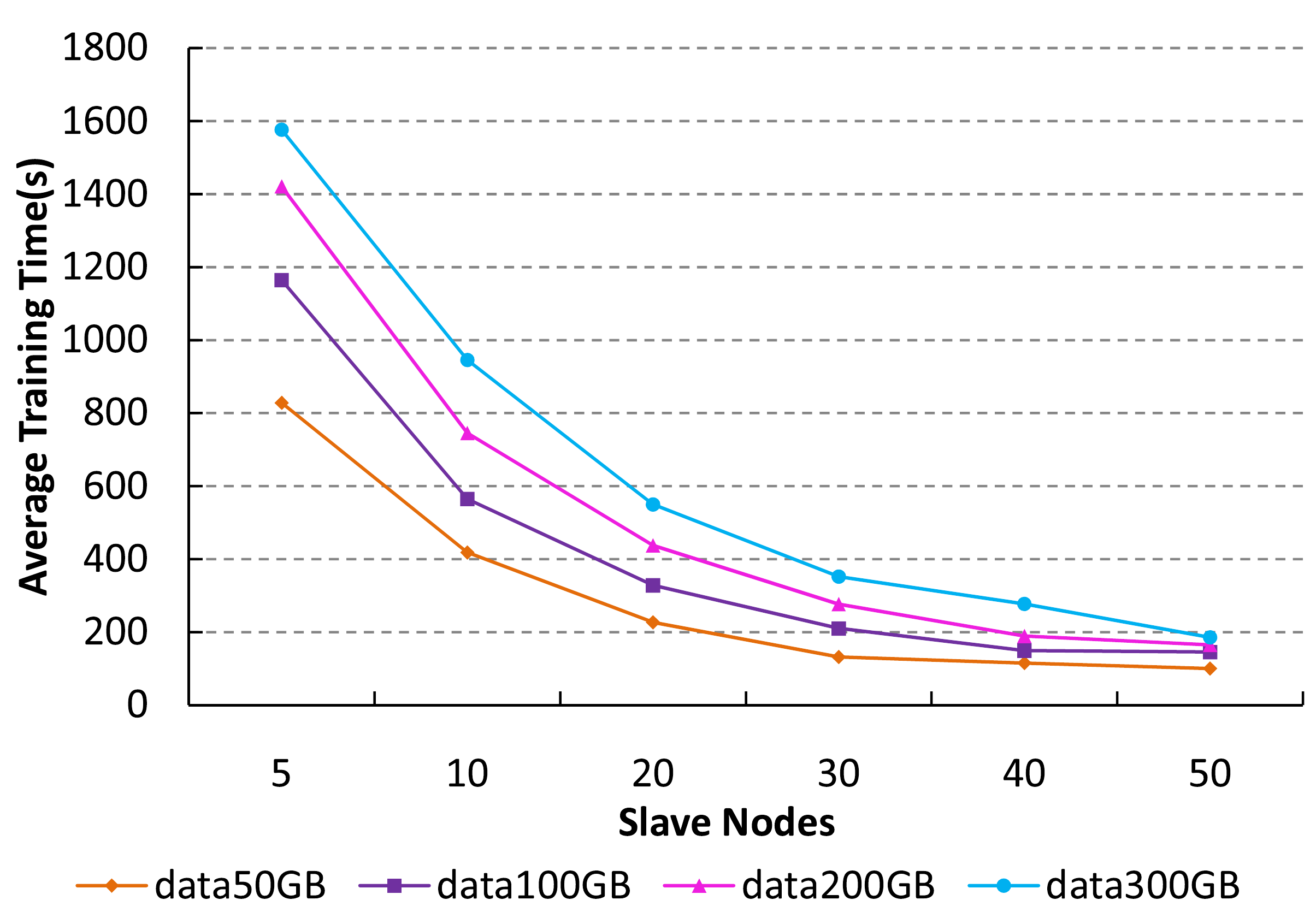}
\caption{Performance evaluation of the PTTP algorithm}
\label{chart09}
\end{figure}

To evaluate the performance of the PTTP algorithm, four groups of historical hospital treatment data are trained at different scales of the Spark cluster.
The sizes of these datasets are 50GB, 100GB, 300GB, and 200GB.
The scale of slave nodes of the Spark cluster in each case increases from 5 to 80.
By observing the average execution time of the PTTP algorithm in each case, different performances across various cases are compared and analyzed.
The results are presented in Fig. \ref{chart09}.

From Fig. \ref{chart09}, the advantage of the parallel algorithm in cases of large-scale data is greater than in cases of small-scale data.
The benefit is more obvious when the number of slave nodes increases.
As the number of cluster nodes increases from 5 to 80, the average execution time of the PTTP model decreases from 879 to 285 s for 300GB of data, and decreases from 328 to 81 s for 50GB of data.

\subsubsection{Performance Evaluation of the HQR System }
\begin{figure}[!ht]
\setlength{\abovecaptionskip}{0pt}
\setlength{\belowcaptionskip}{0pt}
\centering
\includegraphics[width=3.5in]{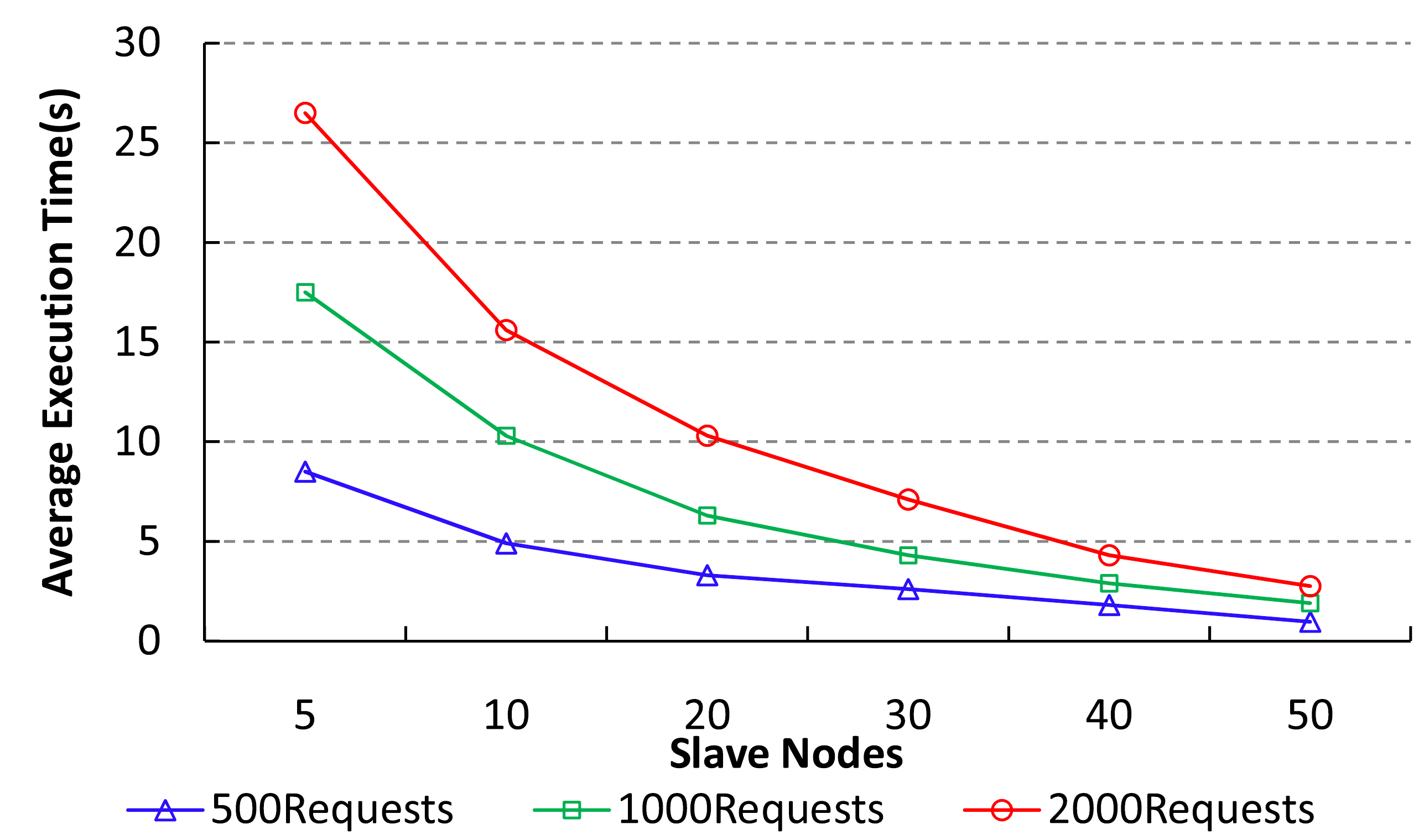}
\caption{Performance evaluation of the HQR system}
\label{chart10}
\end{figure}

The performance of the HQR system is evaluated in this section.
Data for three groups of patients' queuing guidance requirements are executed at the Spark cluster at different scales.
The volumes of requirements data for the recommendation are 500, 1000, and 2000.
The scale of slave nodes of the Spark cluster in each example increases from 5 to 80.
The average execution time of the HQR system for each case is shown in Fig. \ref{chart10}.

In the case of the 5 nodes in the Spark cluster, the average execution time of HQR is 8.5 s for 500 requirements, 17.6 s for 1000, and 26.5 s for 2000.
In the case of 80 nodes in the Spark cluster, the average execution time of HQR is 0.9 s for 500 requirements, 1.9 s for 1000, and 2.7 s for 2000.
As the number of cluster nodes increases from 5 to 80, the average execution times of the HQR system in the three groups decrease at the ratios of 8.85, 9.21 and 9.63 times.
The actual operational results of the algorithm are close to the theoretical results.

\section{Conclusions}
In this paper, a PTTP algorithm based on big data and the Apache Spark cloud environment is proposed.
A random forest optimization algorithm is performed for the PTTP model.
The queue waiting time of each treatment task is predicted based on the trained PTTP model.
A parallel HQR system is developed, and an efficient and convenient treatment plan is recommended for each patient.
Extensive experiments and application results show that our PTTP algorithm and HQR system achieve high precision and performance.

Hospitals' data volumes are increasing every day.
The workload of training the historical data in each set of hospital guide recommendations is expected to be very high, but it need not be.
Consequently, an incremental PTTP algorithm based on streaming data and a more convenient recommendation with minimized path-awareness are suggested for future work.

\section*{Acknowledgment}
This research was partially funded by the Key Program of the National Natural Science Foundation of China (Grant Nos. 61133005, 61432005), the National Natural Science Foundation of China (Grant Nos. 61370095, 61472124), the International Science \& Technology Cooperation Program of China (Grant No. 2015DFA11240), the National Research Foundation of Qatar (NPRP, Grant Nos. 8-519-1-108), and the Natural Science Foundation of Hunan Province of China (Grant Nos. 2016JJ4002).

\ifCLASSOPTIONcaptionsoff
\newpage
\fi

\bibliographystyle{IEEEtran}
\bibliography{reference}

\begin{thebibliography}{10}
\providecommand{\url}[1]{#1}
\csname url@samestyle\endcsname
\providecommand{\newblock}{\relax}
\providecommand{\bibinfo}[2]{#2}
\providecommand{\BIBentrySTDinterwordspacing}{\spaceskip=0pt\relax}
\providecommand{\BIBentryALTinterwordstretchfactor}{4}
\providecommand{\BIBentryALTinterwordspacing}{\spaceskip=\fontdimen2\font plus
\BIBentryALTinterwordstretchfactor\fontdimen3\font minus
  \fontdimen4\font\relax}
\providecommand{\BIBforeignlanguage}[2]{{%
\expandafter\ifx\csname l@#1\endcsname\relax
\typeout{** WARNING: IEEEtran.bst: No hyphenation pattern has been}%
\typeout{** loaded for the language `#1'. Using the pattern for}%
\typeout{** the default language instead.}%
\else
\language=\csname l@#1\endcsname
\fi
#2}}
\providecommand{\BIBdecl}{\relax}
\BIBdecl

\bibitem{ex01}
R.~Fidalgo-Merino and M.~Nunez, ``Self-adaptive induction of regression
  trees,'' \emph{Pattern Analysis and Machine Intelligence, IEEE Transactions
  on}, vol.~33, no.~8, pp. 1659--1672, 2011.

\bibitem{ex02}
S.~Tyree, K.~Q. Weinberger, K.~Agrawal, and J.~Paykin, ``Parallel boosted
  regression trees for web search ranking,'' in \emph{In Proceedings of the
  20th international conference on World wide web(WWW'11)}.\hskip 1em plus
  0.5em minus 0.4em\relax ACM, 2012, pp. 387--396.

\bibitem{ex03}
N.~Salehi-Moghaddami, H.~S. Yazdi, and H.~Poostchi, ``Correlation based
  splitting criterionin multi branch decision tree,'' \emph{Central European
  Journal of Computer Science}, vol.~1, no.~2, pp. 205--220, June 2011.

\bibitem{ex04}
G.~Chrysos, P.~Dagritzikos, I.~Papaefstathiou, and A.~Dollas, ``Hc-cart: A
  parallel system implementation of data mining classification and regression
  tree (cart) algorithm on a multi-fpga system,'' \emph{ACM Transactions on
  Architecture and Code Optimization}, vol.~9, no.~4, pp. 47:1--25, January
  2013.

\bibitem{ex05}
N.~Uyen and T.~Chung, ``A new framework for distributed boosting algorithm,''
  in \emph{Proceeding FGCN '07 Proceedings of the Future Generation
  Communication and Networking}.\hskip 1em plus 0.5em minus 0.4em\relax IEEE,
  2007, pp. 420--423.

\bibitem{ex06}
Y.~Ben-Haim and E.~Tom-Tov, ``A streaming parallel decision tree algorithm,''
  \emph{Journal of Machine Learning Research}, vol.~11, no.~1, p. 849¨C872,
  October 2010.

\bibitem{ex07}
L.~Breiman, ``Random forests,'' \emph{Machine Learning}, vol.~45, no.~1, pp.
  5--32, October 2001.

\bibitem{ex09}
G.~Yu, N.~A. Goussies, J.~Yuan, and Z.~Liu, ``Fast action detection via
  discriminative random forest voting and top-k subvolume search,''
  \emph{Multimedia, IEEE Transactions on}, vol.~13, no.~3, pp. 507 -- 517, June
  2011.

\bibitem{ex10}
C.~Lindner, P.~A. Bromiley, M.~C. Ionita, and T.~F. Cootes, ``Robust and
  accurate shape model matching using random forest regression-voting,''
  \emph{Pattern Analysis and Machine Intelligence, IEEE Transactions on},
  vol.~25, no.~3, pp. 1--14, December 2014.

\bibitem{ex11}
K.~Singh, S.~C. Guntuku, A.~Thakur, and C.~Hota, ``Big data analytics framework
  for peer-to-peer botnet detection using random forests,'' \emph{Information
  Sciences}, vol. 278, pp. 488--497, 2014.

\bibitem{ex12}
S.~Bernard, S.~Adam, and L.~Heutte, ``Dynamic random forests,'' \emph{Pattern
  Recognition Letters}, vol.~33, no.~12, pp. 1580--1586, September 2012.

\bibitem{ex13}
H.~B. Li, W.~Wang, H.~W. Ding, and J.~Dong, ``Trees weighting random forest
  method for classifying high-dimensional noisy data,'' in \emph{IEEE
  International Conference on E-Business Engineering}, vol.~10, November 2010,
  pp. 160--163.

\bibitem{ex08}
G.~Biau, ``Analysis of a random forests model,'' \emph{Journal of Machine
  Learning Research}, vol.~13, no.~1, pp. 1063 -- 1095, January 2012.

\bibitem{ex14}
S.~Meng, W.~Dou, X.~Zhang, and J.~Chen, ``Kasr: A keyword-aware service
  recommendation method on mapreduce for big data applications,''
  \emph{Parallel and Distributed Systems, IEEE Transactions on}, vol.~25,
  no.~12, pp. 3221 -- 3231, 2014.

\bibitem{ex15}
Y.~Y. Chen, A.-J. Cheng, and W.~H. Hsu, ``Travel recommendation by mining
  people attributes and travel group types from community-contributed photos,''
  \emph{Multimedia, IEEE Transactions}, vol.~15, no.~6, pp. 1283--1295, 2013.

\bibitem{ex16}
X.~Yang, Y.~Guo, and Y.~Liu, ``Bayesian-inference based recommendation in
  online social networks,'' \emph{Parallel and Distributed Systems, IEEE
  Transactions on}, vol.~24, no.~4, pp. 642--651, 2013.

\bibitem{ex17}
G.~Adomavicius and Y.~Kwon, ``New recommendation techniques for multicriteria
  rating systems,'' \emph{Intelligent Systems, IEEE}, vol.~22, no.~3, pp.
  48--55, 2007.

\bibitem{ex18}
G.~Adomavicius and A.~Tuzhilin, ``Toward the next generation of recommender
  systems: a survey of the state-of-the-art and possible extensions,''
  \emph{Knowledge and Data Engineering, IEEE Transactions on}, vol.~17, no.~6,
  pp. 734--749, 2005.

\bibitem{ex19}
X.~Wu, X.~Zhu, and G.~Wu, ``Data mining with big data,'' \emph{Knowledge and
  Data Engineering, IEEE Transactions on}, vol.~26, no.~1, pp. 97--107, January
  2014.

\bibitem{ex20}
Apache, ``Hadoop,'' Website, January 2015, \url{http://hadoop.apache.org}.

\bibitem{ex21}
------, ``Spark,'' Website, January 2015, \url{http://spark-project.org}.

\bibitem{ex22}
J.~Dean and S.~Ghemawat, ``Mapreduce: Simplified data processing on large
  clusters,'' \emph{Communications of the ACM}, vol.~51, no.~1, pp. 107--113,
  January 2008.

\bibitem{ex23}
M.~Zaharia, M.~Chowdhury, T.~Das, A.~Dave, J.~Ma, M.~McCauley, M.~J. Franklin,
  S.~Shenker, and I.~Stoica, ``Resilient distributed datasets: A fault-tolerant
  abstraction for in-memory cluster computing,'' in \emph{USENIX NSDI,
  2012}.\hskip 1em plus 0.5em minus 0.4em\relax USENIX, 2012, pp. 1--14.

\bibitem{ex24}
Apache, ``Mahout,'' Website, January 2015, \url{http://mahout.apache.org}.

\bibitem{ex25}
Y.~Xu, K.~Li, L.~He, L.~Zhang, and K.~Li, ``A hybrid chemical reaction
  optimization scheme for task scheduling on heterogeneous computing systems,''
  \emph{IEEE Transactions Parallel Distributed Systems}, vol.~26, no.~12, pp.
  3208--3222, 2015.

\bibitem{ex26}
K.~Li, X.~Tang, B.~Veeravalli, and K.~Li, ``Scheduling precedence constrained
  stochastic tasks on heterogeneous cluster systems,'' \emph{IEEE Transactions
  on Computers}, vol.~64, no.~1, pp. 191--204, January 2015.

\bibitem{ex27}
D.~Dahiphale, R.~Karve, and A.~V. Vasilakos, ``An advanced mapreduce: Cloud
  mapreduce, enhancements and applications,'' \emph{Network and Service
  Management, IEEE Transactions on}, vol.~11, no.~1, pp. 101--115, march 2014.

\bibitem{ex28}
M.~Zaharia, M.~Chowdhury, T.~Das, A.~Dave, J.~Ma, M.~McCauley, M.~J. Franklin,
  S.~Shenker, and I.~Stoica, ``Fast and interactive analytics over hadoop data
  with spark,'' in \emph{USENIX NSDI, 2012}.\hskip 1em plus 0.5em minus
  0.4em\relax USENIX, 2012, pp. 45--51.

\end{thebibliography}

\begin{IEEEbiography}
[{\includegraphics[width=1in,height=1.25in,clip,keepaspectratio]{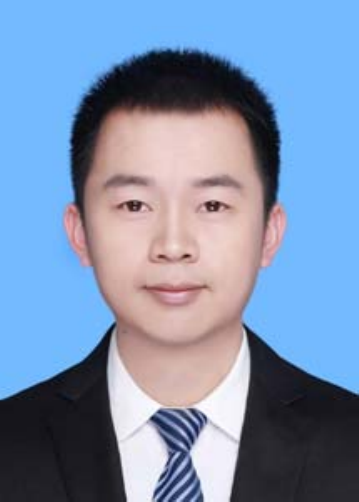}}]
{Jianguo Chen} is currently a Ph.D. candidate in the College of Computer Science and Electronic Engineering at Hunan University, China. His research interests include parallel computing, cloud computing, machine learning, data mining and big data. He has published research articles in international conference and journals of data-mining algorithms and parallel computing.
\end{IEEEbiography}

\begin{IEEEbiography}
[{\includegraphics[width=1in,height=1.25in,clip,keepaspectratio]{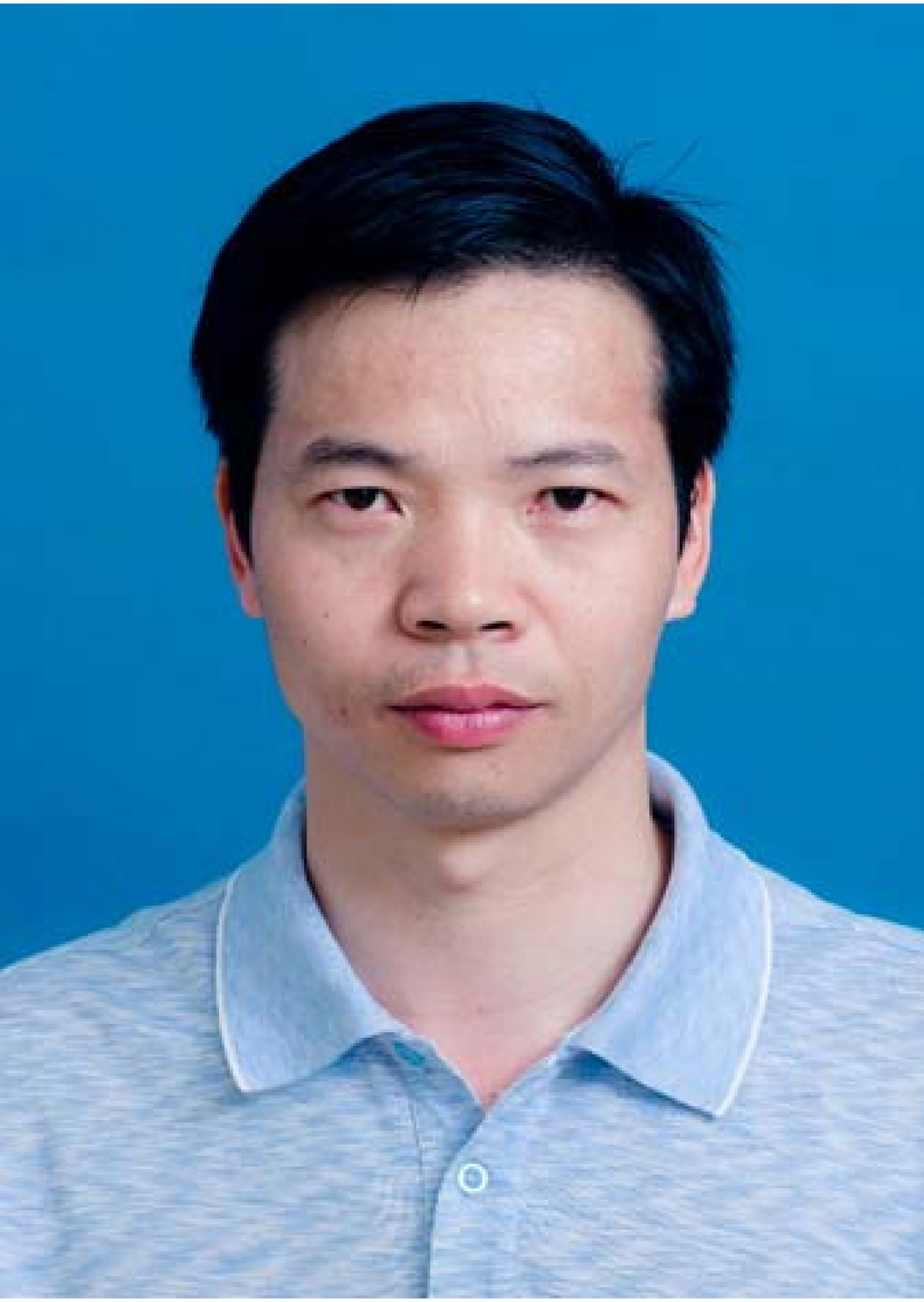}}]
{Kenli Li} (M'11) received the Ph.D. degree in computer science from Huazhong University of Science and Technology, China, in 2003.
He was a visiting scholar at University of Illinois at Urbana-Champaign from 2004 to 2005.
He is currently a full professor of computer science and technology at Hunan University
and deputy director of National Supercomputing Center in Changsha.
His major research areas include parallel computing, high-performance computing, grid and cloud computing.
He has published more than 130 research papers in international conferences and journals, such as IEEE-TC, IEEE-TPDS, IEEE-TSP, JPDC, ICPP, CCGrid.
He is an outstanding member of CCF. He is a member of the IEEE and serves on the editorial board of {\em IEEE Transactions on Computers}.

\end{IEEEbiography}

\begin{IEEEbiography}
[{\includegraphics[width=1in,height=1.25in,clip,keepaspectratio]{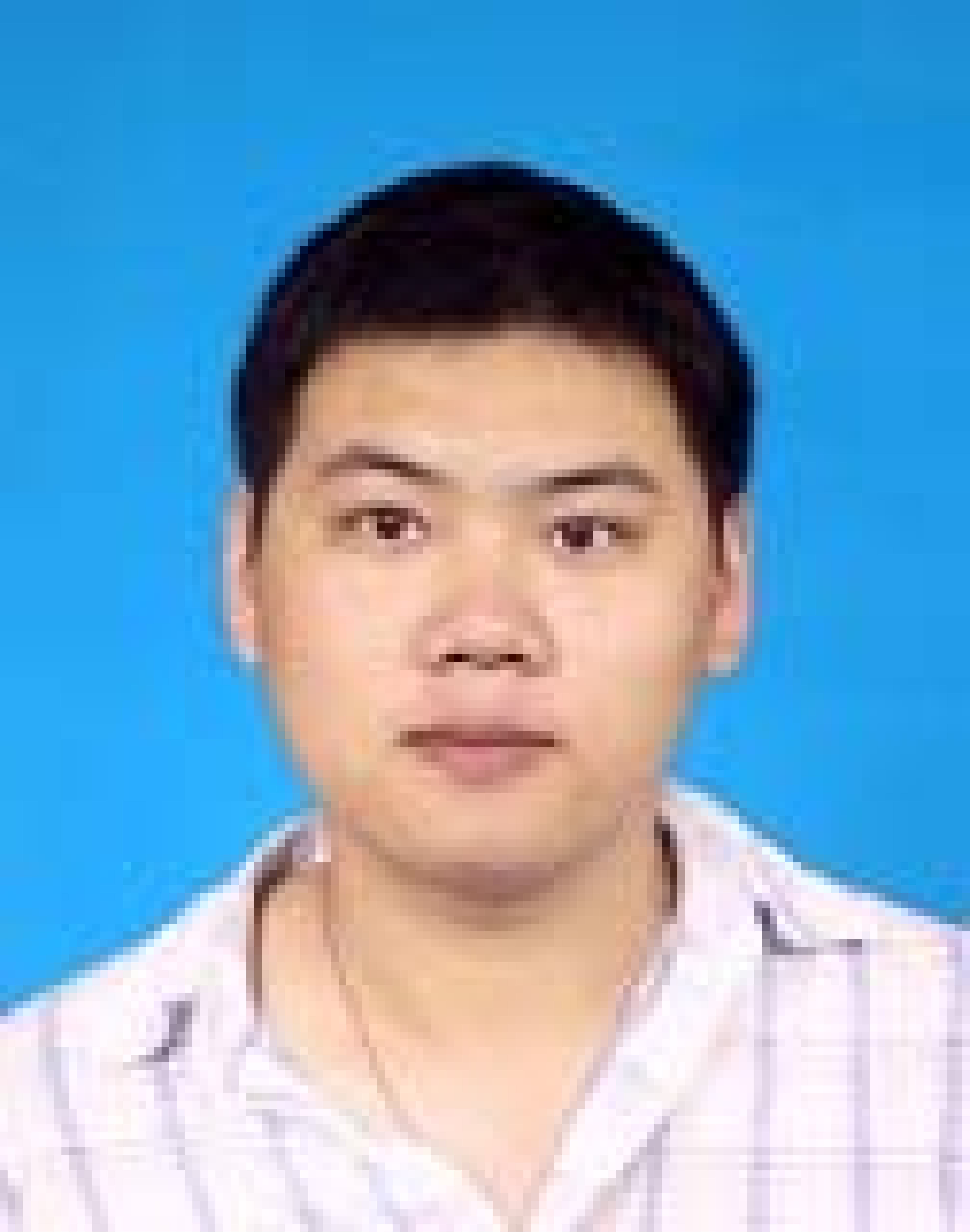}}]
{Zhuo Tang} received the Ph.D. in computer science from Huazhong University of Science and Technology, China, in 2008. He is currently an associate professor of the College of Computer Science and Electronic Engineering at Hunan University, and he is the sub-dean of the department of computing science. His majors are distributed computing system, cloud computing, and the parallel process for big data, include the security model, parallel algorithms, and resources scheduling and management in these areas. He is a member of ACM and CCF.
\end{IEEEbiography}

\begin{IEEEbiography}
[{\includegraphics[width=1in,height=1.25in,clip,keepaspectratio]{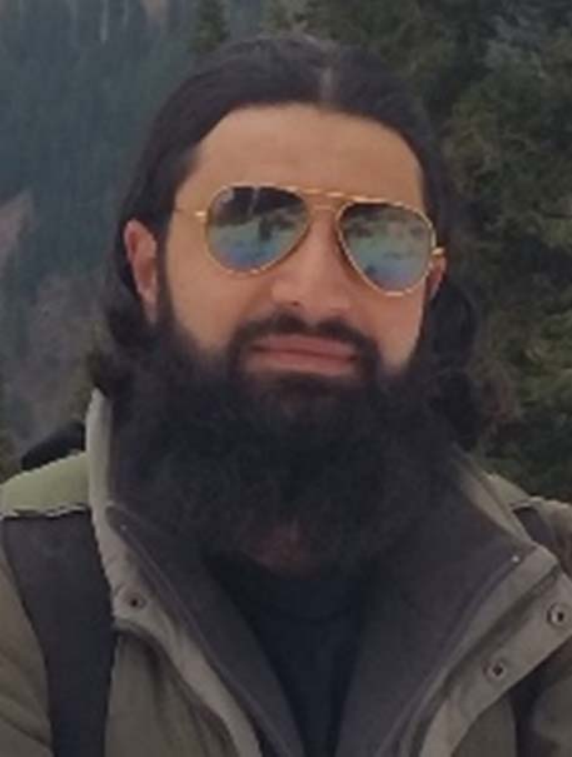}}]
{Kashif Bilal} received his PhD from North Dakota State University USA. He is currently a post-doctoral researcher at Qatar University, Qatar. His research interests include cloud computing, energy efficient high speed networks, and robustness.Kashif is awarded CoE Student Researcher of the year 2014 based on his research contributions during his doctoral studies at North Dakota State University.
\end{IEEEbiography}

\begin{IEEEbiography}
[{\includegraphics[width=1in,height=1.25in,clip,keepaspectratio]{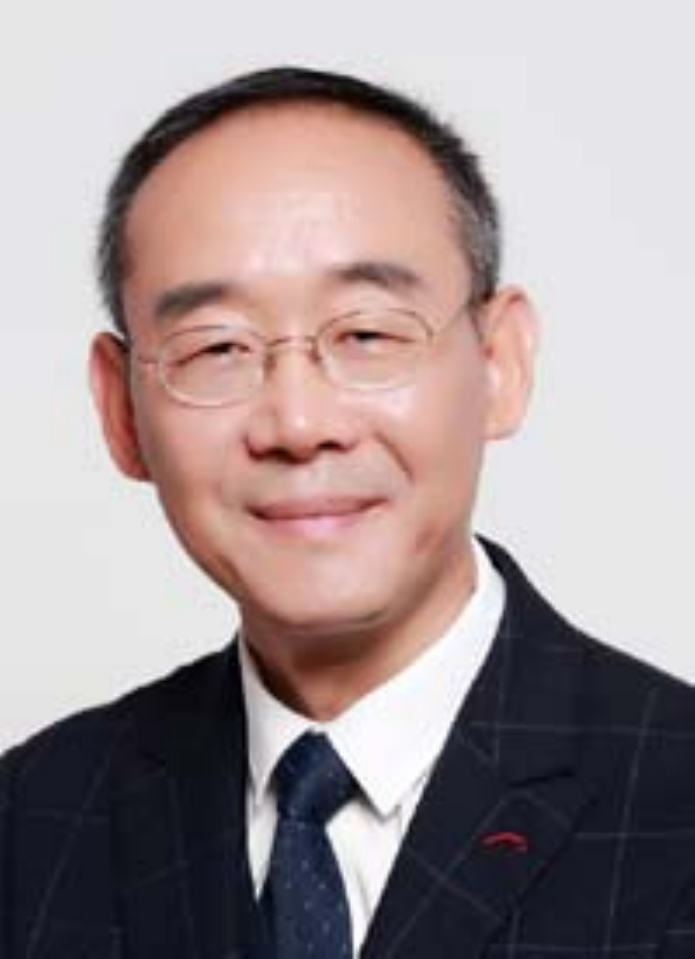}}]
{Keqin Li} (F'15) is a SUNY Distinguished Professor of computer science.
His current research interests include
parallel computing and high-performance computing,
distributed computing,
energy-efficient computing and communication,
heterogeneous computing systems,
cloud computing,
big data computing,
CPU-GPU hybrid and cooperative computing,
multicore computing,
storage and file systems,
wireless communication networks,
sensor networks,
peer-to-peer file sharing systems,
mobile computing,
service computing,
Internet of things and cyber-physical systems.
He has published over 400 journal articles, book chapters, and refereed conference papers,
and has received several best paper awards.
He is currently or has served on the editorial boards of
{\em IEEE Transactions on Parallel and Distributed Systems},
{\em IEEE Transactions on Computers},
{\em IEEE Transactions on Cloud Computing},
{\em Journal of Parallel and Distributed Computing}.
He is an IEEE Fellow.
\end{IEEEbiography}

\end{document}